\documentclass[12pt]{article}
\usepackage{graphicx}
\usepackage{hyperref}

\makeatletter
\def\vereq#1#2{\lower3pt\vbox{\baselineskip1.5pt \lineskip1.5pt
\ialign{$\m@th#1\hfill##\hfil$\crcr#2\crcr\sim\crcr}}}
\makeatother
\newcommand{\als}{\alpha_s}   
\newcommand{\sbar}{\bar{s}}
\newcommand{\gd}{\gamma_\mu}
\newcommand{\gu}{\gamma^\mu}

\newcommand{\su}{\sigma^{\mu\nu}}
\newcommand{\bea}{\begin{eqnarray}}
\newcommand{\eea}{\end{eqnarray}}
\newcommand{\beq}{\begin{equation}}
\newcommand{\eeq}{\end{equation}}
\newcommand{\tm}{\tilde{m}}
\newcommand{\mgluino}{m_{\tilde{g}}}
\begin{document}
\begin{titlepage}
\begin{center}
\today     \hfill %   LBNL-????? \\
~{} \hfill UCB-PTH-02/59  \\
~{} \hfill SLAC-PUB-9612\\
~{} \hfill hep-ph/0212180\\

\renewcommand{\thefootnote}{\fnsymbol{footnote}}

\vskip .1in
{\large \bf Atmospheric Neutrinos Can Make Beauty
Strange}\footnote{The work of RH, DL, and HM
  was supported
  in part by the U.S.  Department of Energy under Contract
  DE-AC03-76SF00098, and in part by the National Science Foundation under
  grant PHY-0098840.  The work of AP was supported by the U.S. Department 
  of Energy under Contract DE-AC03-76SF00515}

\setcounter{footnote}{0}

\vskip 0.3in

Roni Harnik, Daniel T. Larson and Hitoshi Murayama

\vskip 0.05in
{\em Theoretical Physics Group\\
     Ernest Orlando Lawrence Berkeley National Laboratory\\
     University of California, Berkeley, California 94720, USA}

\vskip 0.05in

{\em Department of Physics, University of California\\
     Berkeley, California 94720, USA}

\vskip 0.3in

Aaron Pierce

\vskip 0.05in
{\em Theoretical Physics Group\\
     Stanford Linear Accelerator Center\\
     Stanford University, Stanford, California,94309, USA}
\end{center}

\vskip .1in

 \begin{abstract}
 The large observed mixing angle in atmospheric neutrinos, coupled with
 Grand Unification, motivates the search for a large mixing between
 right-handed strange and bottom squarks.  Such mixing does not appear in
 the standard CKM phenomenology, but may induce significant
 $b \to s$ transitions through gluino diagrams.  Working in the
 mass eigenbasis, we show
 quantitatively that an $\mathcal{O}(1)$ effect on CP violation in
 $B_d^{0} \to \phi K_S$ is possible due to a large mixing between
 $\tilde{s}_R$ and
 $\tilde{b}_R$, while still satisfying constraints from $b \to s
 \gamma$.  We also include the effect of $\tilde{b}_L$-$\tilde{b}_R$
mixing proportional to $m_b\mu\tan\beta$. In the case where $m_{b} \mu
\tan \beta \ll M^2_{SUSY}$ there 
 may be a large effect in $B_s$ mixing correlated with a large
 effect in $B_{d}^{0} \rightarrow \phi K_{S}$, typically 
yielding
% $\Delta M_{B_s} \gtrsim 100$ ps$^{-1}$, 
an unambiguous signal of new physics at Tevatron
 Run II.
\end{abstract}

\end{titlepage}

\newpage

\section{Introduction}
Flavor physics has seen tremendous progress in the past few years.
The discovery of neutrino oscillations by the SuperKamiokande
\cite{Fukuda:1998mi}, SNO \cite{Ahmad:2002jz} , and KamLAND
\cite{Eguchi:2002dm} experiments clearly marks a historic event, while
CP violation has recently been found in two new manifestations: direct
CP violation in the neutral kaon system
\cite{kaon} and indirect CP violation in the the $B_d^{0}$ system
\cite{B}.  On the other hand, we still lack insight into the origin of
flavor and the patterns of masses and mixings.  We need to look for any
possible hints of physics that give us additional insight into these
questions.

One of the major surprises in neutrino physics was the observation
of (two) large angles.  Unlike in the quark sector where all mixing
angles in the Cabibbo--Kobayashi--Maskawa (CKM) matrix are small, both
atmospheric and solar neutrino oscillations require large angles.  An
important question is whether the presence of large angles will give us new
insight into the origin of flavor, masses, and mixings.

It was pointed out in Ref.~\cite{Chang:2002mq}
 that the large angles in the
neutrino sector may imply large angles in the mixing among right-handed
down-type quarks if they are grand-unified with lepton doublets.
Indeed, some $SO(10)$ models with Pati--Salam type unification of
Yukawa matrices suggest that the large mixing angles in neutrinos
arise from the charged lepton mass matrices, and thus also appear in the
down-quark mass matrices.  In these models, one assumes that these 
new large mixing
angles do not appear in the CKM matrix because the right-handed
charged-current interaction is broken at the Pati-Salam unification scale.
However, the imprint of the large atmospheric neutrino mixing angle
may appear in the squark mass matrices as a large
$\tilde{b}_R$-$\tilde{s}_R$ mixing effect though radiative corrections due 
to the large top Yukawa
coupling. The large solar neutrino mixing angle, however, does not cause a 
significant effect because of the smaller Yukawa coupling for lower
generations.  
The new $\tilde{b}_R$-$\tilde{s}_R$ mixing in turn feeds into new effects in 
$B$-physics.  
In particular, there may be large new CP-violating effects in $b
\rightarrow s$ transitions and enhanced $B_s$ mixing.  It has already 
been noted that CP violation in $B_{d}^{0} \to \phi K_S$ is a good place to 
look for new physics effects \cite{Grossman:1996ke,Barbieri:1997kq}.

The time-dependent asymmetry in $B_d^{0} (\overline{B}_d^{0}) \rightarrow \phi
K_S$ was reported recently by both BaBar and BELLE. Their measurements 
differ from the value in the
$J/\psi K_S$ final state by $\mathcal{O}(1)$.  The standard model 
predicts that these 
two channels should give the same value.  The significance of the
difference is $2.7 \sigma$ if the measurements from both collaborations 
are combined-- the current world average for $\sin 2 \beta$ in the 
$B_{d}^{0} \to J/\psi K_{S}$ channel is $0.734 \pm 0.054$, while in 
the $B_{d}^{0} \to \phi K_{S}$ channel $S_{\phi K}=-0.39 \pm 0.41$
\cite{ICHEP02}.    
This report has already sparked many speculations \cite{laundrylist}.  
It is not clear if this is a temporary 
anomaly or a genuine new effect.  Nonetheless it is important to study how 
large the new CP violation in $B_{d}^{0} \to \phi K_S$ can be and how
it is correlated to 
$B_s$ mixing which will be studied soon at Tevatron Run II.

In this paper, we investigate the size of CP violation in $B_{d}^{0}
\rightarrow \phi K_S$ as well as $B_s$ mixing from a potentially large
$\tilde{b}_R$-$\tilde{s}_R$ mixing.  There have also been several
investigations of $B_{d}^{0} \rightarrow \phi K_{S}$ within the context of
supersymmetry (SUSY) \cite{Barbieri:1997kq,Moroi:2000tk,Lunghi:2001af}.  Of the above, only
Ref.~\cite{Lunghi:2001af} investigated the correlation between the
measurement of $\sin 2 \beta$ in $B_{d}^{0} \rightarrow \phi K_{S}$ and
$B_{s}$ mixing.  However it uses the mass insertion formalism, which
is not necessarily appropriate for the large mixing that we will
consider.  In addition, it appeared before the recent experimental
results, and so it did not seek to reproduce such a large shift in
$\sin 2 \beta$.  We perform a calculation in the mass eigenbasis, with a
goal of determining whether supersymmetry can accommodate the central
value of the recent experimental results for $\sin 2 \beta$ in 
$B_{d}^{0} \to \phi K_{S}$.  We then
explore the consequences for $B_{s}$ mixing.

We also emphasize contributions to $\sin 2 \beta$ that arise from a 
combination of $\tilde{b}_{L}$-$\tilde{b}_{R}$ ($m_{b} \mu \tan \beta$) and 
$\tilde{b}_{R}$-$\tilde{s}_{R}$ mixing.  
These contributions, which we find to be important over a wide region of
parameter space, are not easily analyzed in the mass 
insertion approximation.  Analogous combinations were studied in the kaon 
system \cite{kinduced}, but to our knowledge these contributions have not 
been thoroughly analyzed with regard to new physics in the $B_{d}^{0}$ system.

The outline of the paper is as follows. In the next section we
introduce the effective field theory formalism for $b$ decay and work
out $b\rightarrow s$ transitions.  In Section 3 we discuss $B_s$
mixing from large $\tilde{b}_R$-$\tilde{s}_R$ mixing.  Section 4 is
devoted to the discussion of correlations between the $b\rightarrow s$
transition and $B_s$ mixing.  We conclude in Section 5.  Details of
some calculations are presented in the appendices.  In Appendix A
we show the loop functions, while the hadronic matrix elements are estimated 
in Appendix B.

%%%%%%%%%%%%%%%%%%%%%%%%%%%%%%%%%%%%%%%%%%%%%%%%%%%%%%%%%%%%%%%%%%%

\section{CP Violation in $b \rightarrow s$ Transition} \label{sec:bstrans}

In this section we briefly review the well-known effective field theory
formalism for $B$-physics (for a comprehensive review see ~\cite{Buchalla:1995vs}), 
which we 
use to calculate the 
contribution of supersymmetric particles to $b \rightarrow s$ transitions.  
Using this machinery we discuss the 
contribution of a large mixing between right-handed squarks to the 
CP-violating parameter
$S_{\phi K}$.
We also use this formalism to address the constraints on the SUSY 
contribution that come from
the $b \rightarrow s \gamma$ radiative decay.

\subsection{Effective Hamiltonian}

The $b \rightarrow s$ transitions of interest can be described by the 
following $\Delta B=1$ effective Hamiltonian:
\begin{equation}
\mathcal{H}_\mathrm{eff} = \sum_{i=1}^6 \left( C_i \mathcal{O}_i + C'_i
\mathcal{O}'_i \right) + C_\gamma \mathcal{O}_\gamma + C'_\gamma
\mathcal{O}'_\gamma + C_g \mathcal{O}_g + C'_g
\mathcal{O}'_g,
\end{equation}
where
\begin{eqnarray}
\label{eqn:O1}
\mathcal{O}_1 &=& (\sbar_i \gu P_L c_j)(\bar{c}_j \gd P_L b_i) \\
\label{eqn:O2}
\mathcal{O}_2 &=& (\sbar \gu P_L c)(\bar{c} \gd P_L b) \\
\label{eqn:O3}
\mathcal{O}_3 &=& (\sbar \gu P_L b)(\sbar \gd P_L s) \\
\label{eqn:O4}
\mathcal{O}_4 &=& (\sbar_i \gu P_L b_j)(\sbar_j \gd P_L s_i) \\
\label{eqn:O5}
\mathcal{O}_5 &=& (\sbar \gu P_L b)(\sbar \gd P_R s) \\
\label{eqn:O6}
\mathcal{O}_6 &=& (\sbar_i \gu P_L b_j)(\sbar_j \gd P_R s_i) \\
\label{eqn:Ogamma}
\mathcal{O}_{\gamma} &=& \frac{e}{8\pi^2} m_b (\sbar_i \su P_R b_i)
F_{\mu\nu} \\
\label{eqn:Og}
\mathcal{O}_{g} &=& \frac{g_s}{8\pi^2} m_b (\sbar_i \su T^a_{ij}
P_R b_j)G^a_{\mu\nu}. 
\end{eqnarray}
Here $i$ and $j$ are color indices (suppressed in color 
singlet terms), $P_{R,L}=(1\pm \gamma_5)/2$, and $\su=\frac{i}{2}   
[\gamma^\mu,\gamma^\nu]$. The primed operators, which are not    
generated at leading order in the Standard Model, are obtained by taking
$L\leftrightarrow R$ everywhere. Here we have ignored the electroweak
penguin operators $\mathcal{O}_{7-10}$ and the contributions to the  
dipole operators proportional to the $s$-quark mass, $m_s$.   

Following the standard procedure for incorporating QCD corrections  
we match the Wilson coefficients at a high scale to loop diagrams
containing heavy particles present in the full theory, and then use
the renormalization group equations (RGE)
to run the coefficients to the low scale 
where mesons decay. We incorporate leading order QCD corrections using  
the anomalous dimension matrices given in~\cite{Buchalla:1995vs}. The initial
conditions for the standard model coefficients are also
in~\cite{Buchalla:1995vs}; only $C_2$, $C_\gamma$ and $C_g$ are nonzero at
leading order. Leading order running of the standard model 
coefficients has mixing between all eight operators,
$\mathcal{O}_{1-6,\gamma,g}$, due to the presence of a tree level
contribution to $\mathcal{O}_2$. Since right-handed squark mixing only
contributes to the primed operators and gives no tree level contributions,
the leading order SUSY running is simpler: $\mathcal{O}'_{3-6}$ mix
only amongst themselves, as do
$\mathcal{O}'_{\gamma,g}$.\footnote{However, we have checked that even
including mixing between these sets of operators, which is formally at higher
orders in $\alpha_s$, does not affect our
numerical results significantly.}

\begin{figure}[tbp]
  \centering
  \includegraphics[width=0.45\textwidth]{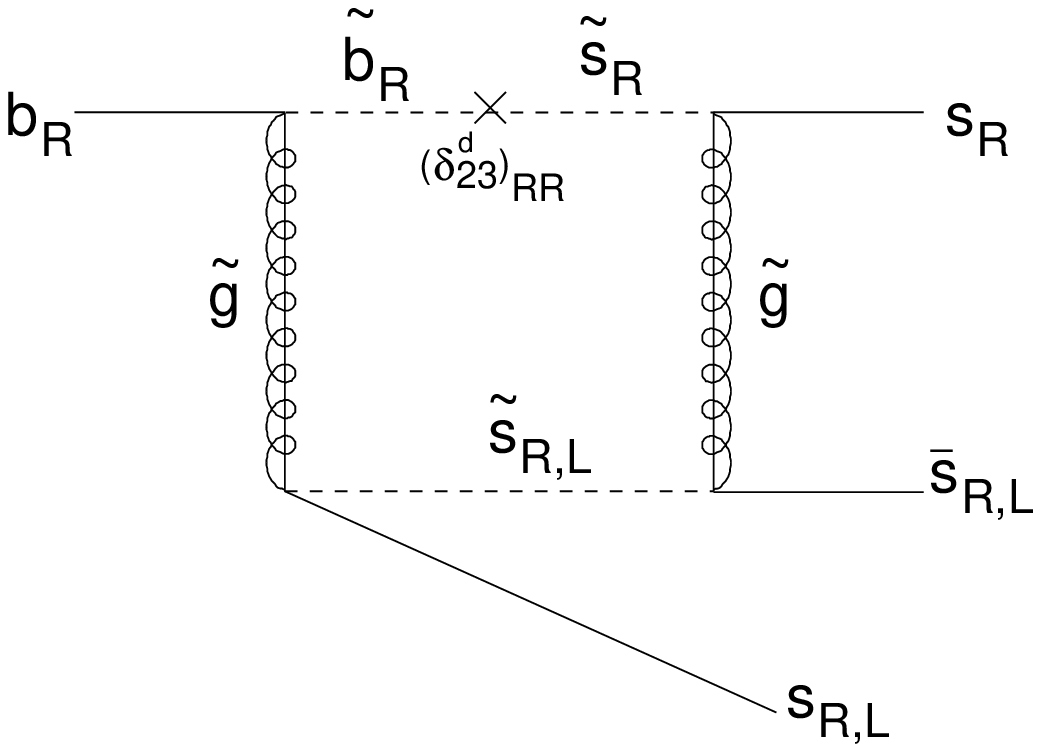}
  \includegraphics[width=0.45\textwidth]{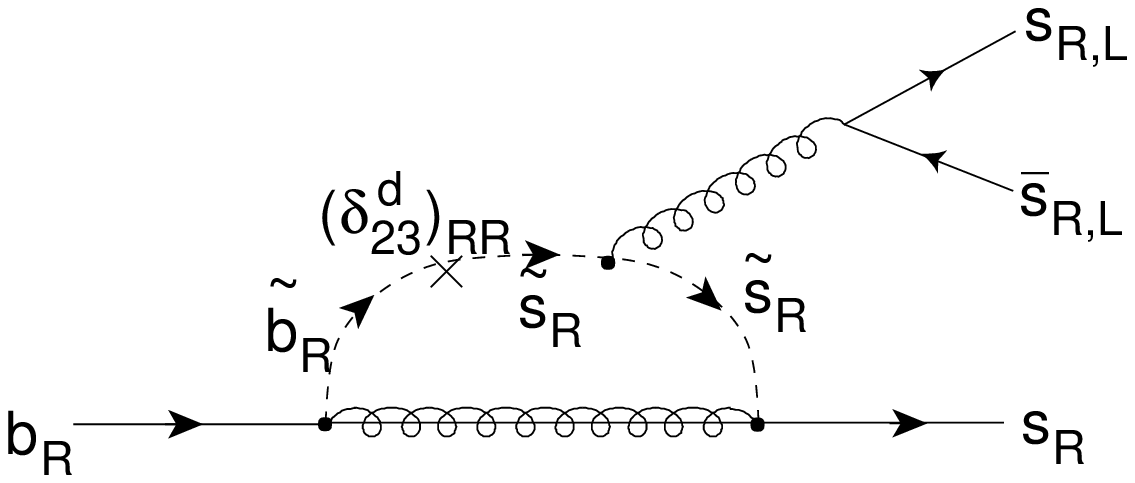}	
  \includegraphics[width=0.45\textwidth]{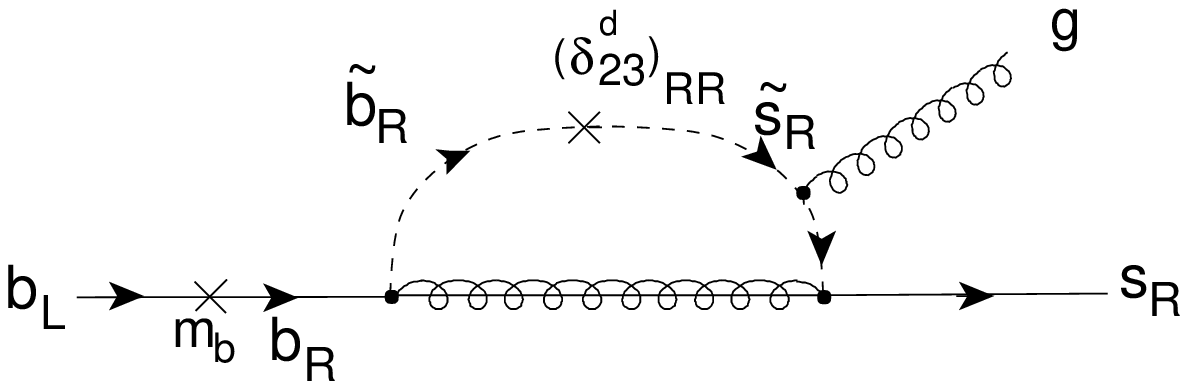}
  \includegraphics[width=0.45\textwidth]{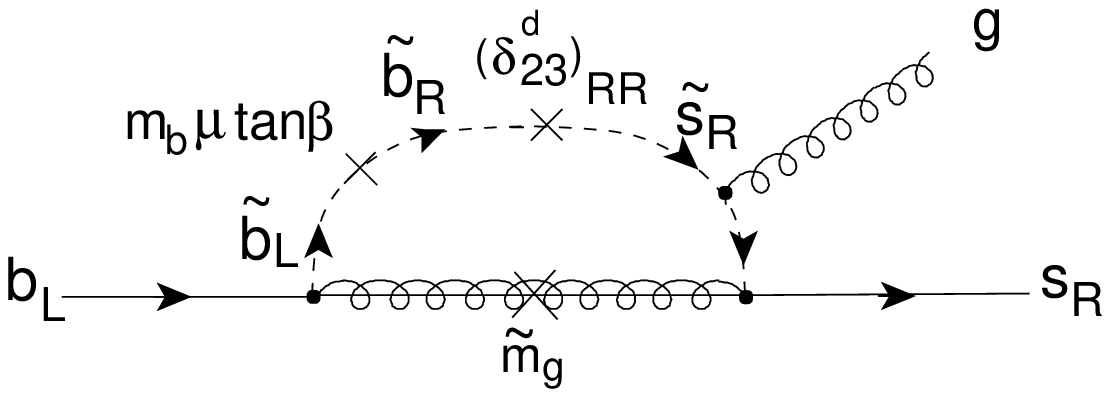}
  \caption{Box and penguin contributions to the $b\rightarrow s\sbar
s$ transition. The bottom row shows contributions to the chromo-dipole
operator.  We show the mass insertions for pedagogical purposes but 
perform calculations in the mass eigenbasis.}
\label{fig:diagrams}
\end{figure}

The SUSY contributions come from box, penguin, and dipole
diagrams. Figure~\ref{fig:diagrams} shows sample diagrams with
mass insertions
schematically indicating the mixing. However, since we are allowing for 
large mixing between the 2nd and 3rd
generation squarks, we use the mass eigenbasis for our
computations. Furthermore, we find that the region where the effect on
$S_{\phi K}$ in the $B_{d}^{0} \rightarrow \phi K_{S}$ channel is
maximized is a region where  
the squarks are highly non-degenerate, again calling into question 
the validity of the mass insertion approximation.\footnote{ Indeed, our
results differ somewhat from previous investigations done in the mass
insertion approximation, e.g. \cite{Lunghi:2001af}.}

The squark mass matrix we consider is motivated by models where a
large right-right  
mixing between the second and third generations is expected, such as
in \cite{Chang:2002mq}
 where this mixing is related to the large mixing in
atmospheric neutrinos.  
In addition to this new contribution we must include other
off-diagonal terms that already exist in the Minimal Supersymmetric 
Standard Model (MSSM), namely
$\tilde{q}_R\tilde{q}_L$ couplings induced by the cross term between
the Yukawa couplings and the $\mu$ term.\footnote{For simplicity we
ignore terms that may arise from trilinear soft terms.} Of the down
type squarks only the third generation can have appreciable
left-right mixing, which is 
proportional to $m_b\mu\tan\beta$. Thus the mass matrix takes the 
approximate form

\begin{equation}
\label{massmatrix}
\tilde{M}^2_{\tilde{d}}= \left(
\begin{array}{cccccc}
\tm_{Ld}^2 & 0 & 0 & 0 & 0 & 0  \\
  0  & \tm_{Ls}^2 & 0 & 0 & 0 & 0\\
  0  & 0 & \tm_{Lb}^2 & 0 & 0 & m_b \mu \tan \beta\\
 0  & 0 & 0 & \tm_{Rd}^2 & 0 & 0\\
 0  & 0 & 0 & 0 &  \tm_{Rs}^2  & \tm_{Rsb}^2\\
 0 & 0& m_b \mu^* \tan \beta  & 0 & \tm_{Rsb}^{*2} & \tm_{Rb}^2 
\end{array} \right).
\end{equation}

We define the mass eigenvalues and mixing matrices as
follows:
\beq
U^\dag \tilde{M}^2_{\tilde{d}} U = \mathrm{diag}(
\tm^2_{L1},\tm^2_{L2},\tm^2_{L3},\tm^2_{R1},\tm^2_{R2},\tm^2_{R3} ),
\eeq
where $U$ is a unitary rotation matrix.  Without loss of generality we 
assume that $\tm_{R3} \leq \tm_{R2}$ (we allow an arbitrary mixing angle).
In the mass eigenbasis $3\times 6$ mixing matrices, $\Gamma^{L,R}$,
appear in the quark-squark-gluino vertices. They are related to $U$ by
\beq
\label{eqn:gammadef}
U_{aA} = \left( \begin{array}{c} \Gamma^{L\ast}_{iA} \\
\Gamma^{R\ast}_{iA} \end{array} \right),
\eeq
where $i=d,s,b$ labels the gauge eigenstates,
$A=1,\ldots,6$ labels the mass eigenstates, and the index $a$ labels
states in the basis $(d_L\ s_L\ b_L\ d_R\ s_R\ b_R)$.
To investigate the effect of 2nd and 3rd generation mixing of the right-handed
squarks we parameterize the mixing matrix as follows:
\beq
\label{eqn:udef}
U =\Phi(\phi_{5},\, \phi_{6}) \, R_{36}(\theta_{36}) \,
R_{35}(\theta_{35}) \, R_{56}(\theta_{56}),
\eeq
where $\Phi = \mathrm{diag}(1,1,1,1,e^{i\phi_5},e^{i\phi_6})$ is a phase
matrix, and $R_{ij}(\theta_{ij})$ is a $2\times 2$ rotation in the
$ij$ plane. The angle $\theta_{35}$ can be solved for using our
assumption that
there is no mixing between $\tilde{b}_L$ and $\tilde{s}_R$.
The Wilson coefficients in the mass eigenbasis were previously given
in~\cite{Moroi:2000tk,Ciuchini:1996vw}.
%and also studied in \cite{Grossman:1999av}.
We reproduce them here, correcting
two typographical errors.
\begin{eqnarray}
\label{eqn:C3}
C'_3 &=& \frac{\als^2}{m_{\tilde{g}}^2} \left( \sum_{AB}
\Gamma^{R\ast}_{sA} \Gamma^{R}_{bA} \Gamma^{R\ast}_{sB}
\Gamma^{R}_{sB} \left[ -\frac{1}{9} B_1(x_A,x_B)-\frac{5}{9}
B_2(x_A,x_B) \right] \right. \nonumber \\ 
&& \ \ \ \ \ \ \ \left. + \sum_A \Gamma^{R\ast}_{sA} \Gamma^{R}_{bA}
\left[ -\frac{1}{18} C_1(x_A)+\frac{1}{2} C_2(x_A) \right] \right) \\
\label{eqn:C4}
C'_4 &=& \frac{\als^2}{m_{\tilde{g}}^2} \left( \sum_{AB}
\Gamma^{R\ast}_{sA} \Gamma^{R}_{bA} \Gamma^{R\ast}_{sB}
\Gamma^{R}_{sB} \left[ -\frac{7}{3} B_1(x_A,x_B)+\frac{1}{3}
B_2(x_A,x_B) \right] \right. \nonumber \\ 
&& \ \ \ \ \ \ \ \left. + \sum_A \Gamma^{R\ast}_{sA} \Gamma^{R}_{bA}
\left[ \frac{1}{6} C_1(x_A)-\frac{3}{2} C_2(x_A) \right] \right) \\
\label{eqn:C5}
C'_5 &=& \frac{\als^2}{m_{\tilde{g}}^2} \left( \sum_{AB}
\Gamma^{R\ast}_{sA} \Gamma^{R}_{bA} \Gamma^{L\ast}_{sB}
\Gamma^{L}_{sB} \left[ \frac{10}{9} B_1(x_A,x_B)+\frac{1}{18}
B_2(x_A,x_B) \right] \right. \nonumber \\ 
&& \ \ \ \ \ \ \ \left. + \sum_A \Gamma^{R\ast}_{sA} \Gamma^{R}_{bA}
\left[ -\frac{1}{18} C_1(x_A)+\frac{1}{2} C_2(x_A) \right] \right) \\
\label{eqn:C6}
C'_6 &=& \frac{\als^2}{m_{\tilde{g}}^2} \left( \sum_{AB}
\Gamma^{R\ast}_{sA} \Gamma^{R}_{bA} \Gamma^{L\ast}_{sB}
\Gamma^{L}_{sB} \left[ -\frac{2}{3} B_1(x_A,x_B)+\frac{7}{6}
B_2(x_A,x_B) \right] \right. \nonumber \\ 
&& \ \ \ \ \ \ \ \left. + \sum_A \Gamma^{R\ast}_{sA} \Gamma^{R}_{bA}
\left[ \frac{1}{6} C_1(x_A)-\frac{3}{2} C_2(x_A) \right] \right) \\
\label{eqn:Cgamma}
C'_\gamma &=& \frac{\als\pi}{m_{\tilde{g}}^2} \left( \sum_A
\Gamma^{R\ast}_{sA} \Gamma^{R}_{bA}
\left[ -\frac{4}{9} D_1(x_A)\right] + \frac{m_{\tilde{g}}}{m_b}
\sum_A \Gamma^{R\ast}_{sA} \Gamma^{L}_{bA}
\left[ -\frac{4}{9} D_2(x_A)\right]\right) \\
\label{eqn:Cglue}
C'_g &=& \frac{\als\pi}{m_{\tilde{g}}^2} \left( \sum_A
\Gamma^{R\ast}_{sA} \Gamma^{R}_{bA} 
\left[ -\frac{1}{6} D_1(x_A)+\frac{3}{2} D_3(x_A) \right]
\right. \nonumber \\
&& \ \ \ \ \ \ \ \left. + 
\frac{m_{\tilde{g}}}{m_b} \sum_A \Gamma^{R\ast}_{sA} \Gamma^{L}_{bA} 
\left[ -\frac{1}{6} D_2(x_A)+\frac{3}{2} D_4(x_A) \right] \right).
\end{eqnarray}
Here we use the definition $x_A = \tm_{A}^2/m^2_{\tilde{g}}$, where
$m_{\tilde{g}}$ is the gluino mass.
The loop functions are given in Appendix~\ref{loopfn}. 
Note that the contributions due to left-right mixing only enter the
dipole operators $C'_\gamma$ and $C'_g$ where they are enhanced by a
factor of $m_{\tilde{g}}/m_b$ over the right-right mixing
contributions to the same operators.
Also notice that with 
our choice of the mixing matrix, $U$, the mass eigenvalues $\tm_{L1}$ and
$\tm_{R1}$ do not enter the Wilson coefficients. 
Because there was some disagreement in the literature, we have explicitly 
recomputed the box contributions (proportional to $B_1$ and
$B_2$). However, the 
penguin contributions (proportional to $C_{1,2}$ and $D_{1-4}$) 
are well established. See for example,~\cite{Gabbiani:1996hi}.
We found several inconsistencies in the literature which can be remedied as
follows. In Equation~(A.8) of~\cite{Moroi:2000tk} the coefficient
$-\frac{1}{4N_c}$ of $B_1$ should be replaced by $-\frac{1}{N_c}$ and
the expression for $\tilde{C}^\mathrm{DM}_R$ in Equation (A.16) should
be multiplied by $-\frac{i}{2}$. In Equation~(42) of~\cite{Lunghi:2001af} the
factor of $\frac{7}{18}$ should be $\frac{7}{6}$, and in Equation~(43)
the factor of $\frac{16}{9}$ should be
$\frac{8}{9}$. The loop function $M_2(x)$ in~\cite{Gabbiani:1996hi} should be
multiplied by $-x$ instead of $-\frac{1}{x}$. Finally,
in~\cite{Barbieri:1997kq} there are
typos in each line of Equations (B.4a-e) and in (B.6a,b).

\subsection{$B_{d}^{0} \rightarrow \phi K_S$}
We now specialize our discussion to the $B_{d}^{0} \rightarrow \phi K_S$ decay, 
with the goal of computing the contribution to the CP asymmetry measured in 
this channel.  
In addition to the Wilson coefficients 
we have presented, we must also compute the hadronic matrix elements of 
the operators.  
The calculation of these matrix elements is non-perturbative, so 
approximations must be made in order for us to make progress.
In the naive factorization
approximation we break each matrix element up into a
pair of color singlet currents, one which creates the $\phi$ from the
vacuum and the other that mediates the $B_{d}^{0} \rightarrow K$
decay, and we discard any color-octet currents.

For the
operators $\mathcal{O}^{(\prime)}_{3-6}$ there are two
ways of contracting the external quarks with the quark fields in the
operator. After employing Fierz transformations as necessary, and 
using the identity
$\delta_{ij}\delta_{k\ell} = \frac{1}{N}\delta_{i\ell}\delta_{kj} + 2
T^a_{i\ell} T^a_{kj}$ to rearrange color indices to form singlet
currents, we arrive at the following matrix elements~\cite{Barbieri:1997kq}:
\begin{eqnarray}
\langle \phi K_S|\mathcal{O}_{3,4}|\bar{B}^0_d\rangle &=& \frac{1}{4}
H \left( 1+\frac{1}{N_c}\right) \\
\langle \phi K_S|\mathcal{O}_5|\bar{B}^0_d\rangle &=& \frac{1}{4} H \\
\langle \phi K_S|\mathcal{O}_6|\bar{B}^0_d\rangle &=& \frac{1}{4} H
\frac{1}{N_c}
\end{eqnarray}
where $H=2(\epsilon_\phi\cdot p_B) f_\phi m^2_\phi
F_+(m_\phi^2)$ (see Appendix~\ref{matrix-element} for definitions of
the decay constant and form factors). 
The same results hold for the
matrix elements of the corresponding primed
operators because the axial vector currents do not contribute, so
the chirality of the operators is irrelevant. We take $N_c=3$
throughout our analysis.

The matrix elements of the chromo-dipole operators 
$\mathcal{O}^{(\prime)}_g$ are more
difficult to analyze, so we show the details explicitly in
Appendix~\ref{matrix-element} following~\cite{Barbieri:1997kq}.
These manipulations yield
\begin{equation}
\label{eqn:kappadef}
\langle \phi K_S|\mathcal{O}_{g4}|\bar{B}^0_d\rangle = \kappa
\frac{\als}{2\pi} H \frac{N_c^2-1}{2N_c^2}. \\
\end{equation}
where our definition of $\kappa$ agrees
with~\cite{Moroi:2000tk} up to a sign convention.\footnote{Note that
in~\cite{Moroi:2000tk} the overall
factor of $\frac{1}{4}$ in the first line of Equation~(18) should not be
multiplying the last term that contains $\kappa_\mathrm{DM}$.}
Numerically we find $\kappa =
-\frac{9}{8} + \mathcal{O}\left( \frac{m_\phi^2}{m_B^2} \right) 
\simeq -1.1$. However, there were many assumptions about the quark
momenta that go into this estimation of $\kappa$, so the numerical
value of $-1.1$ should be taken as a guideline only. We will present our
results for various values of $\kappa$ to demonstrate the dependence.

There is one final ingredient in the Standard Model contribution to
the amplitude.  This comes from the one-loop matrix element of
$\mathcal{O}_2$ when
the charm quarks are closed into a loop. 
It is given by~\cite{Deshpande:1994pw} $P=
\frac{\als}{8\pi} C_2 (\frac{10}{9} + G(m_c,\mu,q^2))$ with
\begin{equation}
G(m,\mu,q^2)=4\int_0^1 dx\,x(1-x) \ln\frac{m^2-x(1-x)q^2}{\mu^2}.
\end{equation}
Numerically we use $m_c=1.35$ GeV and $q^2=\frac{m_b^2}{2}$ which gives
$P=(-0.015-0.011 i)C_2$.

Putting the factorized matrix elements together the amplitude for
$\bar{B}_d \rightarrow \phi K_S$ can be written
\begin{eqnarray}
\label{eqn:bphikshad}
\bar{\mathcal{A}}^\mathrm{SM}_{\phi K} & = & H \left[ \frac{1}{4} \left(
1+\frac{1}{N_c}\right) (C_3 + C_4) + \frac{1}{4}
C_5 +\frac{1}{4}\frac{1}{N_c} C_6 \right] \nonumber \\
& + &  H \left[\frac{2}{9} P + \kappa
\frac{\als}{2\pi}\frac{N_c^2-1}{2N_c^2} C_{g}   \right]
\end{eqnarray}
\begin{eqnarray}
\bar{\mathcal{A}}^\mathrm{SUSY}_{\phi K} & = &  H \left[ \frac{1}{4} \left(
1+\frac{1}{N_c}\right) (C'_3 + C'_4) + \frac{1}{4}
C'_5 +\frac{1}{4}\frac{1}{N_c} C'_6 \right] \nonumber \\ 
& +  & H \left[\kappa
\frac{\als}{2\pi}\frac{N_c^2-1}{2N_c^2} C'_{g}   \right]
\end{eqnarray}

The time-dependent $CP$-asymmetry is given by
\begin{equation}
a_{\phi K}(t) = C_{\phi K} \cos(\Delta M_{B_d^{0}} t) + S_{\phi K}
\sin(\Delta M_{B_d^{0}}t),
\end{equation}
where
\begin{equation}
C_{\phi K} = \frac{1-|\lambda|^2}{1+|\lambda|^2} \,, \ \ \ \ \ \ \ \ \ \
S_{\phi K} = \frac{2\,\mathrm{Im} \lambda}{1+|\lambda|^2} \; .
\end{equation}
Here $\lambda$ is defined as
\begin{equation}
\lambda = \frac{q}{p}\frac{\mathcal{A}(\bar{B}^0_d\rightarrow \phi
K_S)} {\mathcal{A}(B^0_d\rightarrow \phi K_S)} =
\frac{q}{p}\frac{(\bar{\mathcal{A}}^\mathrm{SM}_{\phi K} +
\bar{\mathcal{A}}^\mathrm{SUSY}_{\phi K})} {(\mathcal{A}^\mathrm{SM}_{\phi
K}+\mathcal{A}^\mathrm{SUSY}_{\phi K})} \; . 
\end{equation}
The ratio $q/p$ from $B_{d}^{0}$ mixing is dominated by the standard model
and is nearly a pure phase, $e^{i2\beta}$, where $\beta$ is the standard
angle of the unitarity triangle. In the Standard Model the ratio of
amplitudes $\bar{\mathcal{A}}/\mathcal{A}$ is real, i.e. there is 
no CP violation in the decay, rather all CP violation results from
mixing. On the other hand,
$\mathcal{O}(1)$ phases in the supersymmetric contribution can give 
the ratio a phase, $\phi^\mathrm{SUSY}$. Then we have
\begin{equation}
\lambda = e^{i(2\beta+\phi^\mathrm{SUSY})}
\frac{|\bar{\mathcal{A}}|}{|\mathcal{A}|} \ \ \ \Rightarrow S_{\phi
K} = \sin (2\beta+\phi^\mathrm{SUSY}).
\end{equation}
Thus the presence of a phase in the down squark mixing matrix can alter the
measured value of $S_{\phi K}$ from the standard model prediction of
$S_{\phi K}=\sin 2\beta_{J/\psi K}=0.73$. The amount of deviation is
described in Section \ref{numerical}.

Also note that $P$ possesses a strong phase that is not present
in $\mathcal{A}^\mathrm{SUSY}$. The presence of
a weak phase in $\mathcal{A}^\mathrm{SUSY}$ then allows
for the possibility of nonzero direct $CP$-violation, namely
$C_{\phi K}\neq 0$. We do not pursue this signature further here, 
as quantitative statements are difficult due to the large hadronic 
uncertainties.

\subsection{Constraints from $b\rightarrow s \gamma$}
\begin{figure}[tbp]
  \centering
  \includegraphics[width=0.45\textwidth]{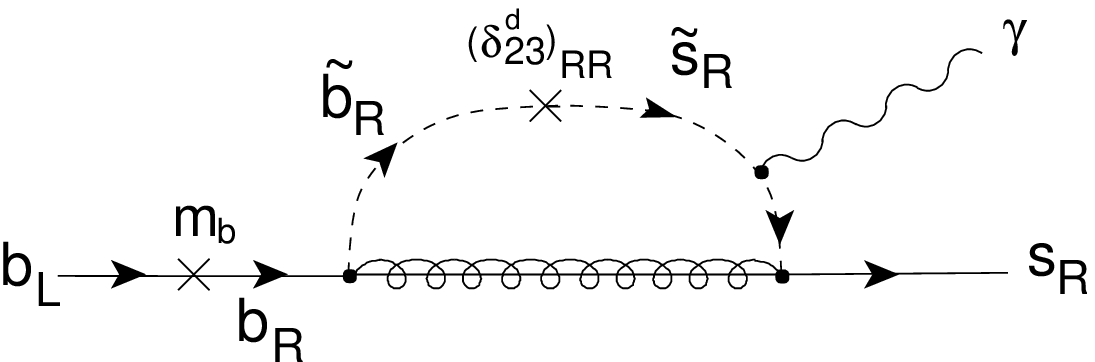}
  \includegraphics[width=0.45\textwidth]{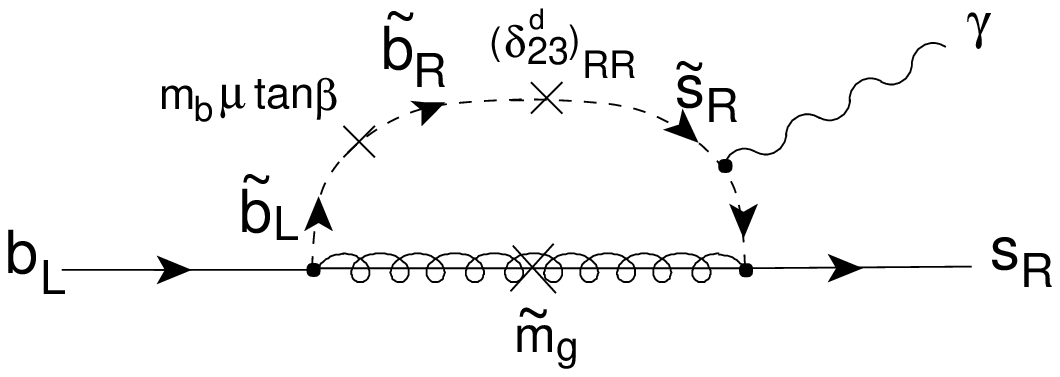}
  \caption{Sample contributions to the $b\rightarrow s \gamma$ transition.  
	We show the mass insertions here for pedagogical purposes,
	but perform calculations in the mass eigenbasis.}
\label{fig:b2sgamma}
\end{figure}

A large mixing between right-handed strange and bottom squarks generates 
the operator $\mathcal{O}'_\gamma$ through penguin
diagrams, as in Figure~\ref{fig:b2sgamma}. Therefore the
tight experimental constraints on the branching ratio 
$BR(B\rightarrow X_s \gamma)$ serve to
limit the contributions from squark mixing.  

In the model we consider there are two important 
contributions to the $\mathcal{O}'_\gamma$ operator.  We can classify the
contributions according to where the helicity flip for 
$\mathcal{O}'_{\gamma}$ arises.  
In the first contribution, the helicity flip is present on the external 
$b$-quark line, and gives a contribution proportional to the $b$-mass.
This contribution is present even when the only mixing between the squarks 
is an off-diagonal mixing between the right-handed squarks of the second and 
third generation.  The constraint on this contribution is relatively
mild.\footnote{We note that the limits of reference~\cite{Gabbiani:1996hi} assume 
that 
new contributions to $b \to s \gamma$ are summed incoherently.  In general, 
this will underestimate the contribution of the $LL$ mass insertion.}   
The second contribution has a helicity flip on the gluino line, so
is enhanced relative to the first contribution by a factor of 
$\mgluino/m_{b}$  (see Equations~(\ref{eqn:Cgamma}) and~(\ref{eqn:Cglue})).  
This contribution is only present if there is left-right mixing in the 
squark matrix.  Because of the $\mgluino/m_{b}$ 
enhancement, this contribution is relatively strongly constrained.  In our 
framework, this contribution arises only from the combination of a left-right 
mixing between the $\tilde{b}$ squarks and the right-right mixing between the 
$\tilde{b}$ and $\tilde{s}$ squarks. The result is that for large 
values of $\mu \tan \beta$, a smaller 
$\tilde{b}_{R}$-$\tilde{s}_{R}$ mixing is allowed.
 
When there is no significant off-diagonal mixing among the left-handed
squarks, we can write:
$BR(b\rightarrow s \gamma) \propto |C_\gamma|^2 + |C'_\gamma|^2$, 
where the first contribution is from the standard model and the 
second is from supersymmetric
penguins.\footnote{The two contributions are added incoherently
because they contribute to different final helicity states of the
$s$-quark.} 

A recent theoretical evaluation within the Standard Model gives 
\cite{Gambino:2001ew}:
\begin{equation}
BR(b \rightarrow s \gamma)_{th}=(3.60 \pm .30) \times 10^{-4}.
\end{equation}
After rescaling to limit the photon energies to $E _{\gamma} > 1.6$ GeV 
(for details see \cite{Gambino:2001ew}),
an averaging of experimental results from BaBar, BELLE, CLEO, and 
ALEPH~\cite{bsgexpts} yields
\begin{equation}
BR(b \rightarrow s \gamma)_{exp}=(3.29\pm .34) \times 10^{-4}.
\end{equation}
The experimentally measured branching ratio is actually slightly 
smaller than the standard model prediction, which leaves little room for 
new physics contributions. Subtracting experiment from theory we find:
\begin{equation}
BR(b \rightarrow s \gamma)_{th}- BR(b \rightarrow s \gamma)_{exp}=(.31
\pm .45) \times 10^{-4}.
\end{equation}
We will require that the supersymmetric contribution keep the 
theoretical prediction within $2\sigma$ of the experimentally
measured value.  This means additional contributions from supersymmetry 
can be roughly $1/6$ of those in the Standard Model.
For simplicity, and to avoid the
theoretical uncertainty associated with the direct calculation of the
branching ratio, we will constrain the supersymmetric contributions by 
requiring $|C'_\gamma|^2 \leq 0.16  \times|C_\gamma|^2$ where both 
coefficients are calculated to leading order. Thus we are making the 
simplifying assumption that the higher order QCD corrections affect the two
operators in the same way.

\subsection{Numerical Analysis}
\label{numerical} 

Within the framework we have chosen, motivated by atmospheric neutrino
oscillations, there are four mass eigenvalues, two mixing angles, and
two phases in the down-squark mass matrix that enter in the
computations of $S_{\phi K}$ and $BR(b\rightarrow s\gamma)$. 
However, the fact that the neutron electric dipole moment (EDM) has not been observed strongly
constrains the phase of $\mu$, especially for large $\mu\tan\beta$. 
We have checked that allowing a non-zero phase of $\mu$ does not substantially
affect even our quantitative conclusions.  Therefore we conservatively
take the phase of $\mu$ to be zero for the remainder of this paper.
Including the gluino mass we are then left with eight essentially unknown 
parameters. In order to reduce the size of the parameter space we will 
investigate two limiting cases: the case where the (flavor-diagonal) 
mixing between 
left- and right-handed $\tilde{b}$ squarks is negligible and the case 
where such 
mixing, when coupled with the large right-right mixing, leads to the dominant 
contribution.  We refer to the latter case as ``$LR+RR$ mixing.''

Note that by taking $\mu$ to be real, the remaining phase 
only appears as an overall phase in $\Gamma_{R}$.  As a result, all 
SUSY diagrams have the same phase.

\subsubsection{Dominant Right-Right Mixing}
\label{rrmixing}

First we consider the situation where the contribution from
$\tilde{b}_L$-$\tilde{b}_R$
mixing is negligible, i.e. $m_b\mu\tan\beta \ll \tilde{m}^2$, the
mass scale of the squarks.\footnote{Within 
the MSSM $\mu \tan \beta$ cannot go to zero while satisfying 
experimental constraints. We find, however, that there is a 
portion of parameter space above the smallest experimentally  
allowed value of $\mu \tan \beta$ where the right-right mixing diagrams are 
dominant. 
Furthermore, right-right mixing dominates when $\tm_R\ll\tm_L$.} 
Here the parameter space is reduced:
in this limit there is only one mixing angle, one phase, and three mass 
eigenvalues that enter the computation of the Wilson coefficients. 
The presence of mixing with an order one phase in the right-handed
down squark sector can significantly alter the measured value of
$S_{\phi K}$.  Our first question is whether a large right-right mixing 
between the down squarks can reproduce the central value for 
$\sin 2 \beta$ in the $\phi K_{S}$ channel measured at the $B$-factories.  
We find that using the central value of our estimate for $\kappa=-1.1$, 
it is possible to
reproduce the observed central value and accommodate the constraints 
from $b \rightarrow s \gamma$.  However, this estimate for $\kappa$ is 
highly uncertain, and increasing the magnitude of $\kappa$ increases the 
contribution to $S_{\phi K}$ without changing the contribution 
to $b \rightarrow s \gamma$.  Therefore, we present our numerical results 
for two cases,  $\kappa=-1.1$, and a value with greater magnitude, 
$\kappa=-2$, which we still view as reasonable given the substantial 
uncertainties involved in its estimation.

\begin{figure}[htbp]
  \centering
  \includegraphics[width=0.49\textwidth]{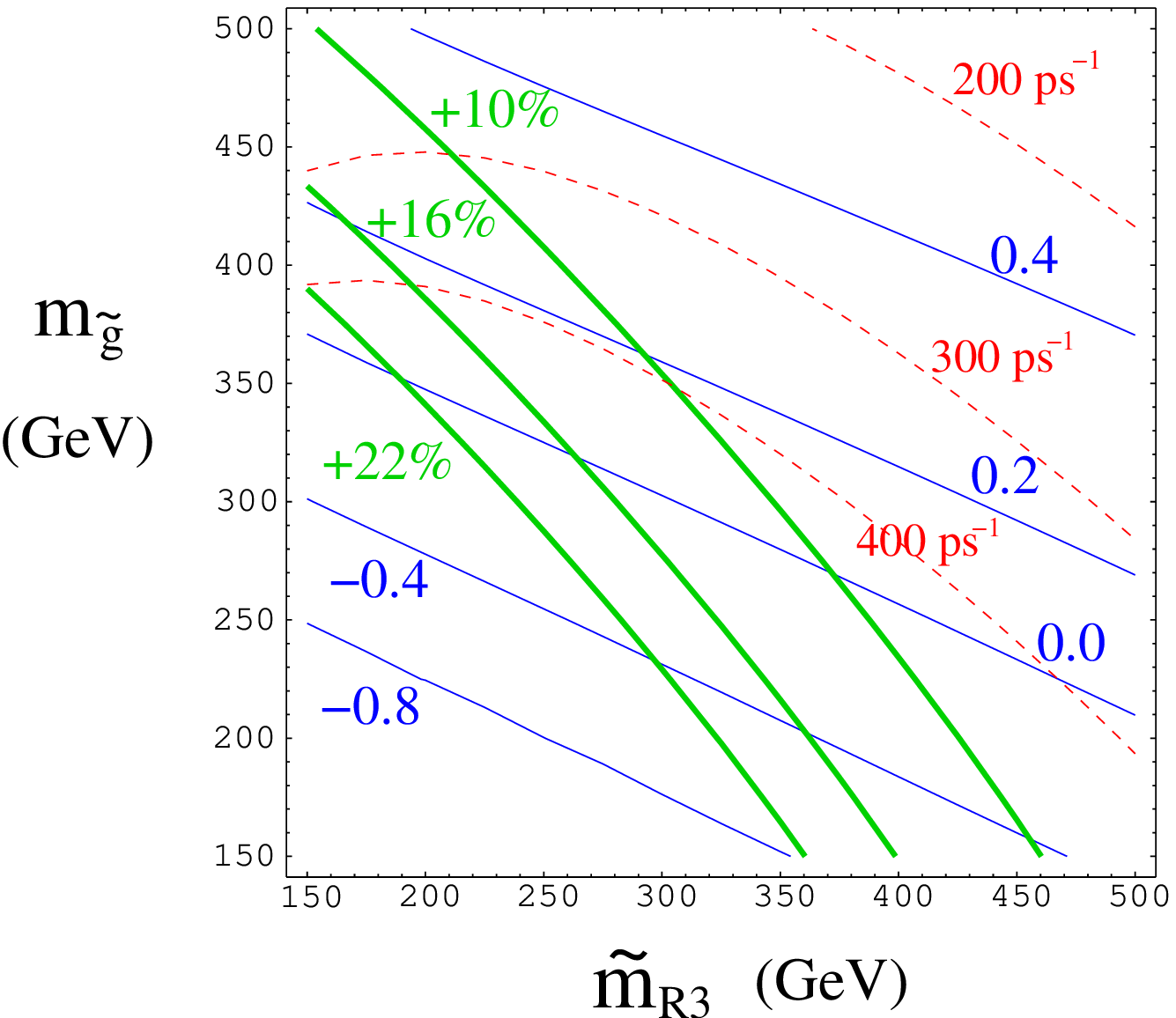}
  \includegraphics[width=0.49\textwidth]{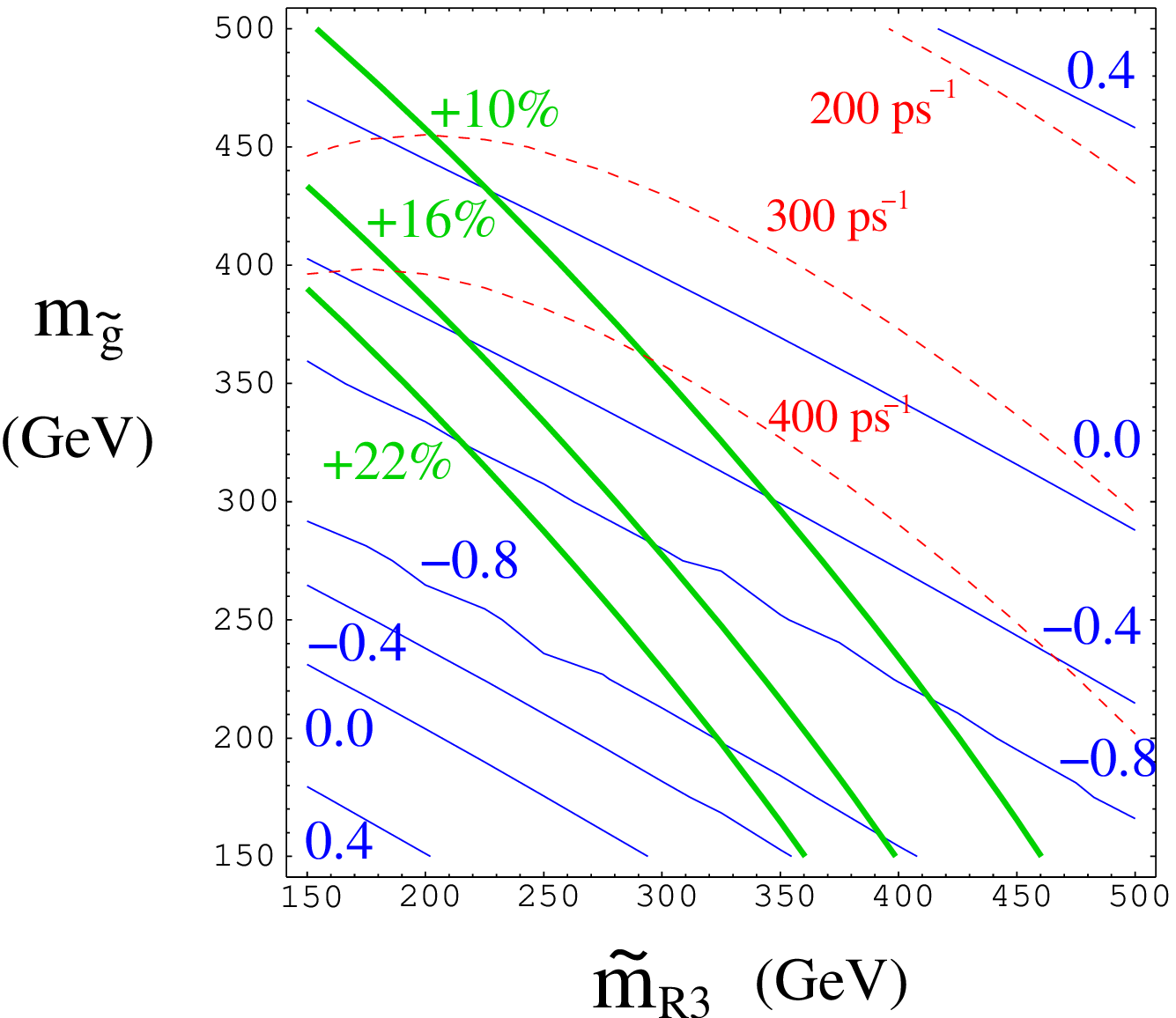}
  \caption{Contours of $S_{\phi K}$ (thin blue), percent increase in 
  $BR(b\rightarrow s \gamma)$ (thick green), and $\Delta
  M_{B_s}$ (dashed red) for $\tm_{R2} = 5$ TeV, optimal mixing angle
  and phase, and $\kappa=-1.1$ (left) and $\kappa=-2.0$ (right).}
\label{fig:contour1}
\end{figure}

In Figure~\ref{fig:contour1} we show contours of $S_{\phi K}$
as a function of the gluino mass and $\tm_{R3}$. 
We have also chosen values for the mixing angle and phase
in $\Gamma^{R}$ which give the greatest deviation of $S_{\phi K}$ from
the Standard Model prediction. Also shown are contours of the percent
increase in $BR(b\rightarrow s \gamma)$ due to new physics and the
corresponding values of $\Delta M_{B_S}$ (the latter will be discussed
in Section \ref{bsmixing}). For gluino masses around $200$ GeV, $S_{\phi
K}$ can take on values as low as $-0.4$ for $\kappa=-1.1$ while still
keeping the increase in $BR(b\rightarrow s \gamma)$ below 16\%. For
$\kappa=-2.0$ the value of $S_{\phi K}$ can reach all the way to $-1.0$.

Generally speaking, lighter squark and gluino masses increase the
effect of the new physics contributions, allowing $S_{\phi K}$ to
depart from the standard model expectation. But at the same time this
increases the
contribution to $BR(b\rightarrow s \gamma)$ and runs up against the
experimental constraint.

\begin{figure}[htbp]
  \centering
  \includegraphics[width=0.49\textwidth]{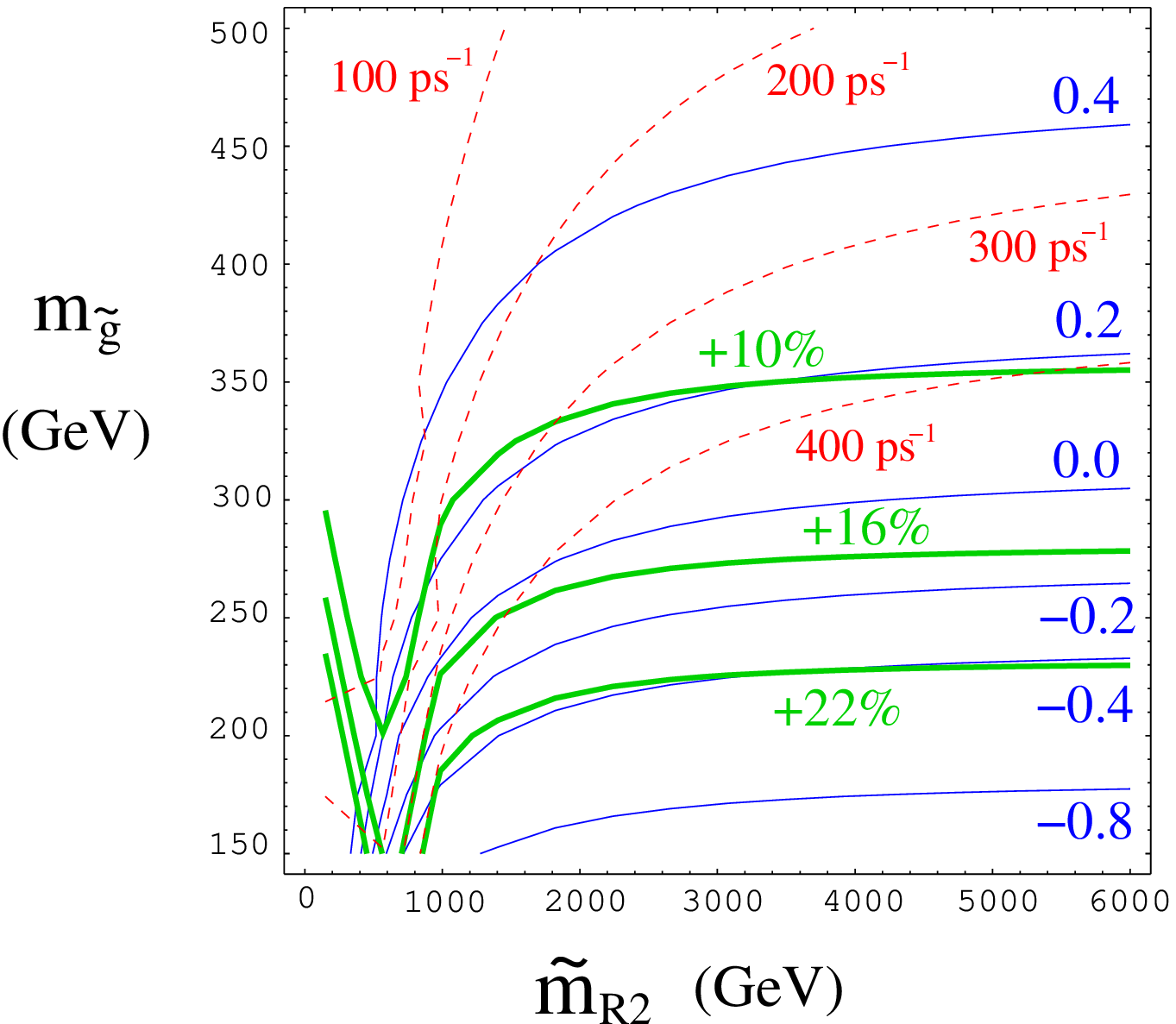}
  \includegraphics[width=0.49\textwidth]{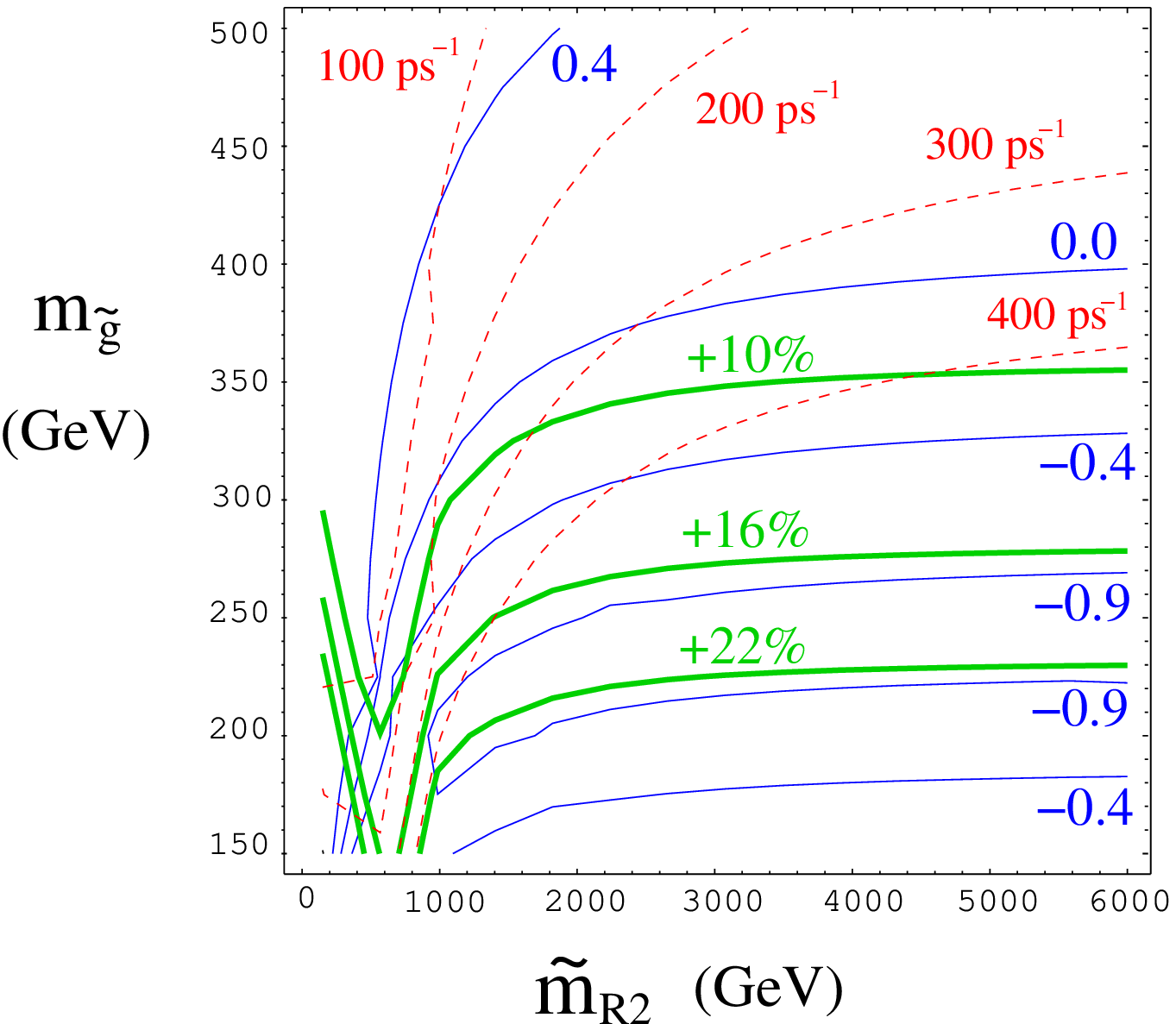}
  \caption{Contours of $S_{\phi K}$ (thin blue), percent increase in 
  $BR(b\rightarrow s \gamma)$ (thick green), and $\Delta
  M_{B_s}$ (dashed red) for $\tm_{R3} = 300$ GeV, optimal mixing
  angle and phase, and $\kappa=-1.1$ (left) and $\kappa=-2.0$
  (right).}
\label{fig:contour2}
\end{figure}

In Figure~\ref{fig:contour2} we plot the same contours as a function of
the gluino mass and the heavier squark mass $\tm_{R2}$, with
$\tm_{R3}= 300$ GeV. For $\kappa=-1.1$ there is a range of gluino
masses where  $S_{\phi K}$ can be below zero while 
%the increase in $BR(b\rightarrow s \gamma)$ is less than 16\%.
still satisfying the $b\rightarrow s \gamma$ constraint.  
The minimum possible
$S_{\phi K}$ allowed by this constraint decreases as the magnitude of
$\kappa$ increases. For example, with $\kappa=-2.0$ the minimum value
of $S_{\phi K}$ is roughly $-0.8$.

\begin{figure}[htbp]
  \centering
  \includegraphics[width=0.45\textwidth]{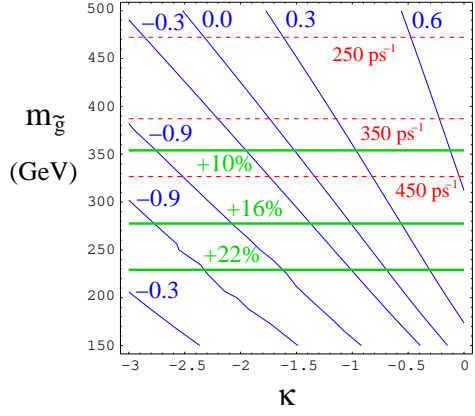}
  \caption{Contours of $S_{\phi K}$ (thin blue), percent increase in 
  $BR(b\rightarrow s \gamma)$ (thick green), and $\Delta
  M_{B_s}$ (dashed red) for $\tm_{R2}=5$ TeV, $\tm_{R3} = 300$ GeV, and
  optimal mixing angle and phase.}
\label{fig:contour3}
\end{figure}

Finally in Figure~\ref{fig:contour3} we plot the same contours as a
function of the gluino mass and $\kappa$ with $\tm_{R2}=5$ TeV
and $\tm_{R3}=300$ GeV.
The $b\rightarrow s \gamma$ constraint is independent of
$\kappa$, and we see how $S_{\phi K}$
decreases with the increasing magnitude of $\kappa$.


\subsubsection{Dominant $LR+RR$ Mixing}
\label{lrmixing}

Now we consider a second limiting case, where the diagonal left-right mixing
leads to the dominant contribution to the $b$-quark decays. 
The contribution from $LR+RR$ mixing is enhanced in the dipole operators, so we
may focus our attention on the coefficients $C'_g$ and $C'_\gamma$. To
evade the constraint from $b \rightarrow s \gamma$, while simultaneously
getting a large effect in $S_{\phi K}$, we want to minimize
the ratio $C'_\gamma/C'_g$, which can be done by taking larger values
of $x$, i.e. squark masses much heavier than the gluino.

\begin{figure}[tbp]
  \centering
  \includegraphics[width=0.49\textwidth]{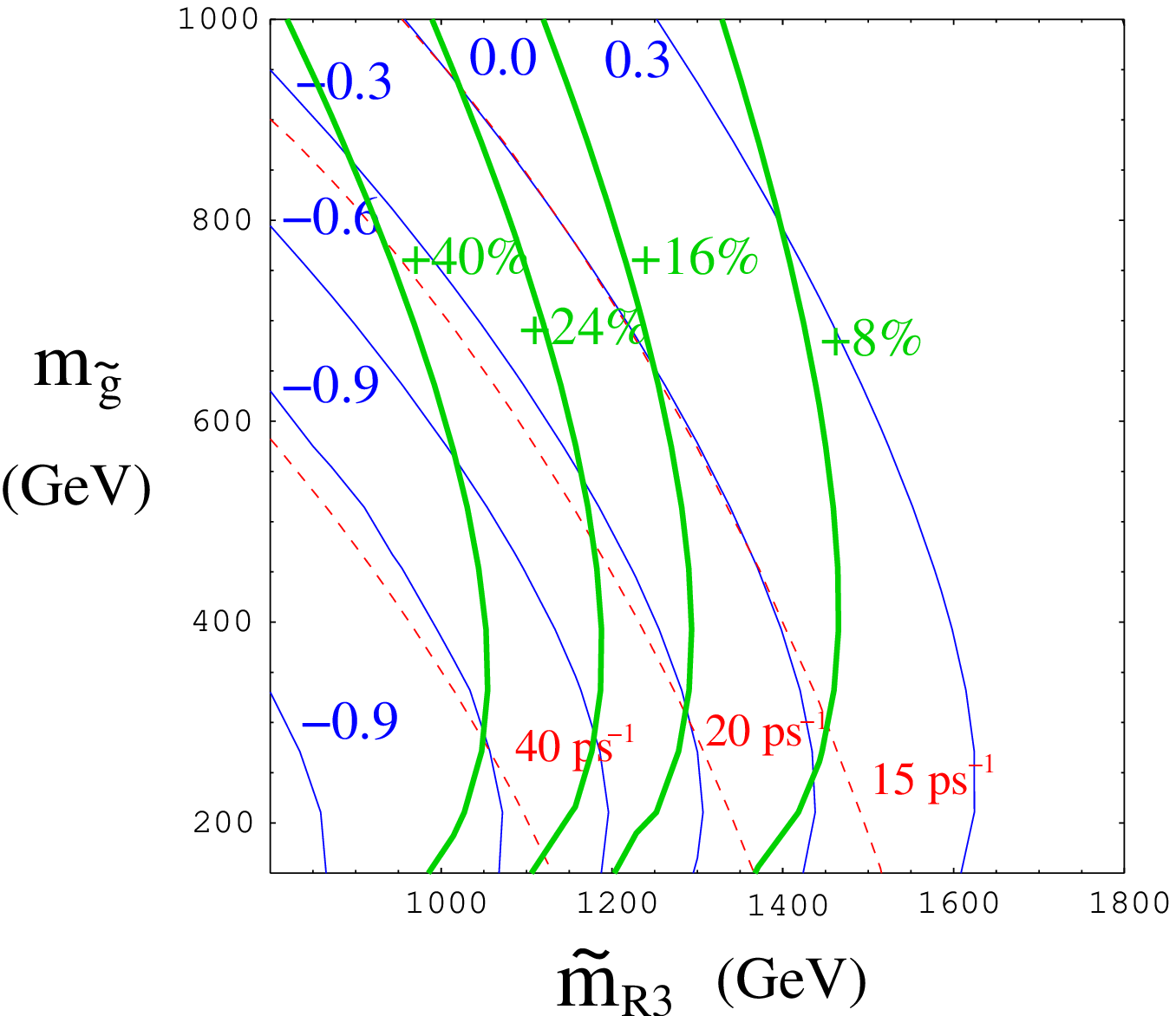}
  \includegraphics[width=0.48\textwidth]{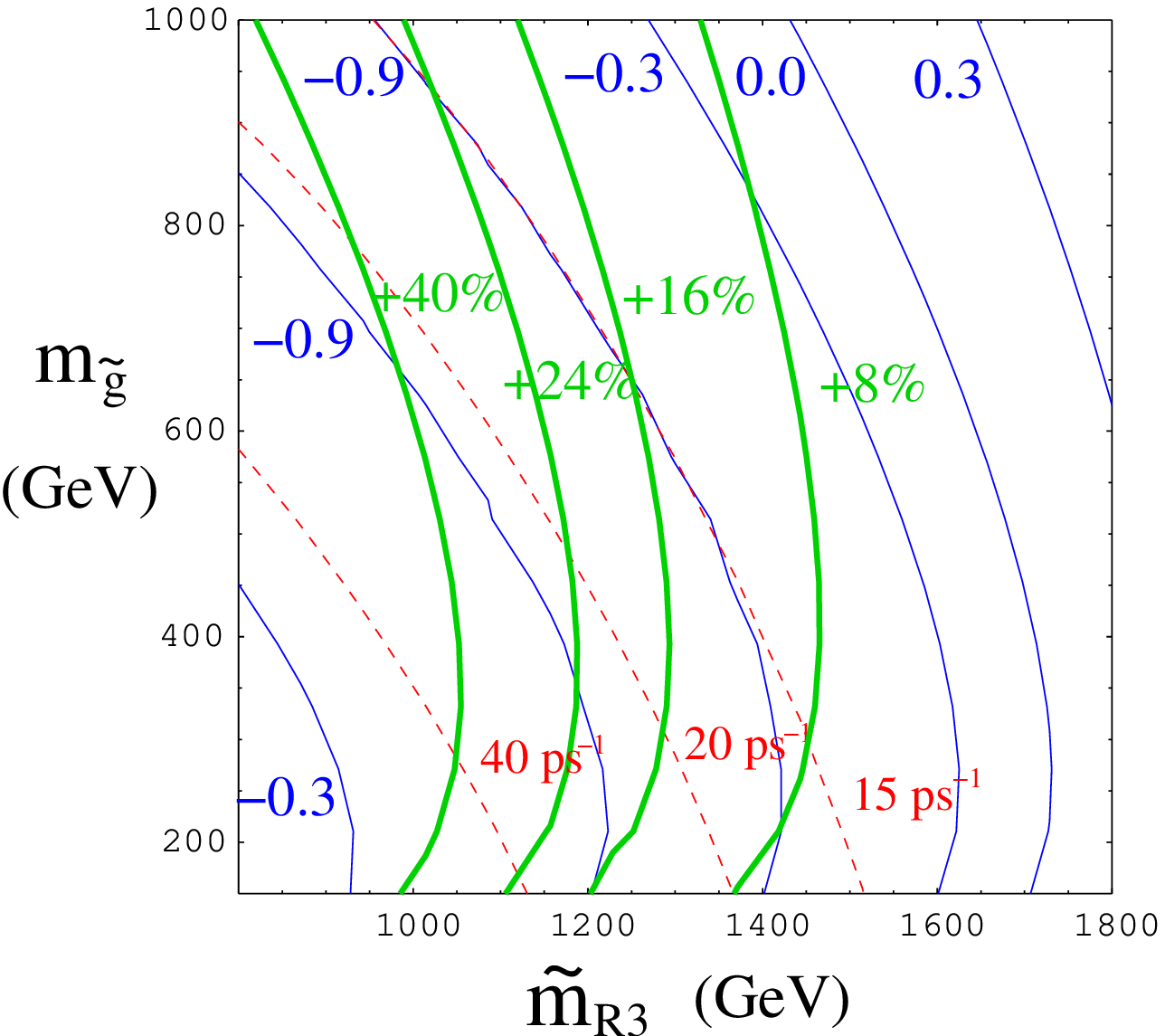}
  \caption{Contours of $S_{\phi K}$ (thin blue), percent increase in 
  $BR(b\rightarrow s \gamma)$ (thick green), and $\Delta
  M_{B_s}$ (dashed red) for $\tm_{L2}, \, \tm_{L3}, \, \tm_{R2} \sim 2$ TeV, 
  optimal mixing angles and phases, $\mu\tan\beta \sim 35$ TeV, 
  and $\kappa=-1.1$ (left) and $\kappa=-2.0$ (right).}
\label{fig:lrcontour1}
\end{figure}

In Figure~\ref{fig:lrcontour1} we reproduce
Figure~\ref{fig:contour1} but with a large value of $\mu\tan\beta
\sim 35$ TeV. In this region the lightest squark mass eigenvalue, 
$\tm_{R3}$, needs to be above 1 TeV to avoid the bound from
$b\rightarrow s\gamma$. For $\kappa=-1.1$ the smallest $S_{\phi K}$
can be is about $-0.4$ for a small gluino mass, while for
$\kappa=-2.0$ any value is possible. In this case the $\kappa$-dependence
is very simple because the main 
contribution to $S_{\phi K}$ comes from a
single operator $\mathcal{O}'_g$ whose contribution is directly
proportional to $\kappa$. Thus an increased absolute value of $\kappa$
directly increases the effect in $S_{\phi K}$ without affecting the
bound from $b\rightarrow s\gamma$.

\subsubsection{Combination of Contributions}
\begin{figure}[tbp]
  \centering
  \includegraphics[width=0.7\textwidth]{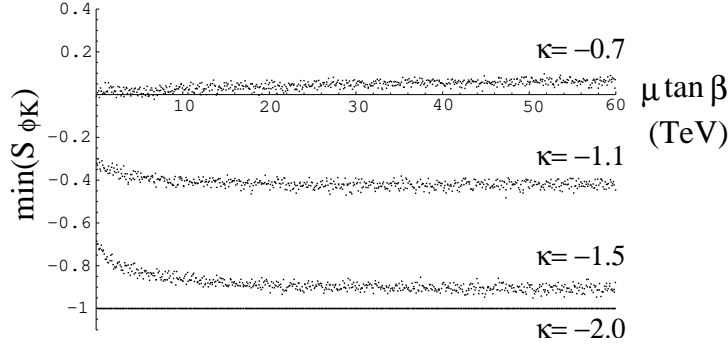}
  \caption{Minimum values of $S_{\phi K}$ as a function of
$\mu\tan\beta$ for various values of $\kappa$ resulting from a scan of
  58 million points that satisfied
the $b\rightarrow s\gamma$ constraint. Note that $\kappa=-2.0$ can give
 $S_{\phi K} = -1.0$ for all values of $\mu\tan\beta$.}
\label{fig:mins2b}
\end{figure}

In order to ascertain the relevance of each of these two regimes we
scanned over the parameter space searching for the minimal values of
$S_{\phi K}$ as a function of the product $\mu\tan\beta$. The result
is shown in Figure~\ref{fig:mins2b} for sevaral values of
$\kappa$. From the figure it is clear that for $\kappa=-1.1$ there is
a slightly larger effect on $S_{\phi K}$ for larger values of
$\mu\tan\beta$, though the entire region allows for a substantial
deviation from the standard model. Notice that once $\kappa$ reaches
$-2.0$ any value of $S_{\phi K}$ is possible.
%As expected, increasing the magnitude of
%$\kappa$ increases the effect on $S_{\phi K}$ somewhat more for the
%larger values of $\mu\tan\beta$ where the chromo-dipole operator gives the
%main contribution. 

To get a sense for the relative size
of the $RR$ and $LR+RR$ contributions to $S_{\phi K}$, we can compare the
magnitude of the two terms comprising $C^\prime_g$, Equation~(\ref{eqn:Cglue}).
Not surprisingly, for $\mu\tan\beta$ greater than about 25 TeV the $LR+RR$
contribution dominates by an order of magnitude. However, even for
$\mu\tan\beta < 1$ TeV there can be points where the $LR+RR$ contribution
to the chromo-dipole operator is
just as important as the $RR$ contribution.  This underscores the importance 
of treating this calculation in the mass eigenbasis.

\section{$B_s$ Mixing}
\label{bsmixing}

\begin{figure}[tp]
  \centering
  \includegraphics[width=0.45\textwidth]{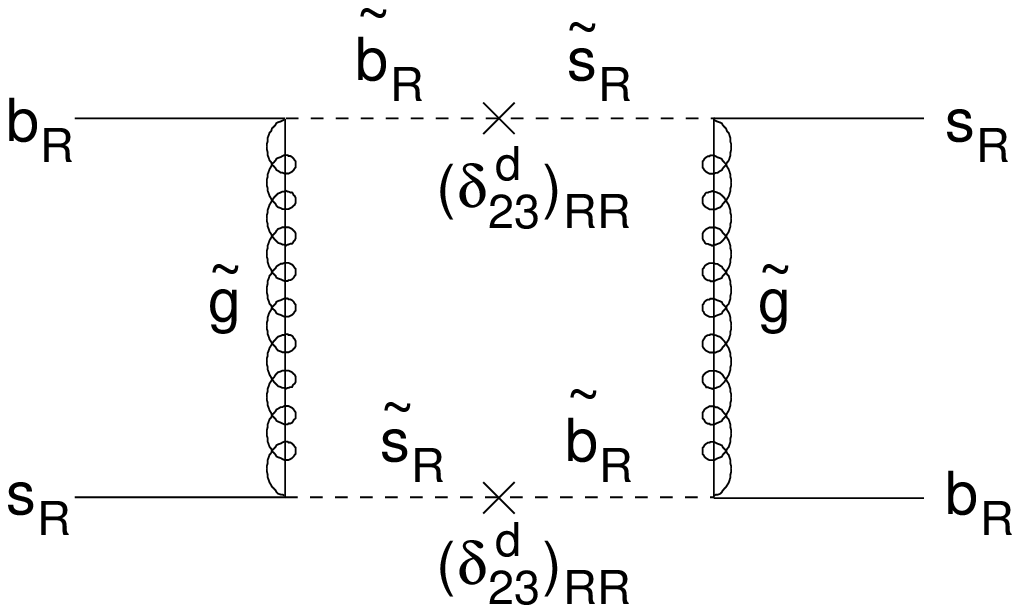}
  \includegraphics[width=0.45\textwidth]{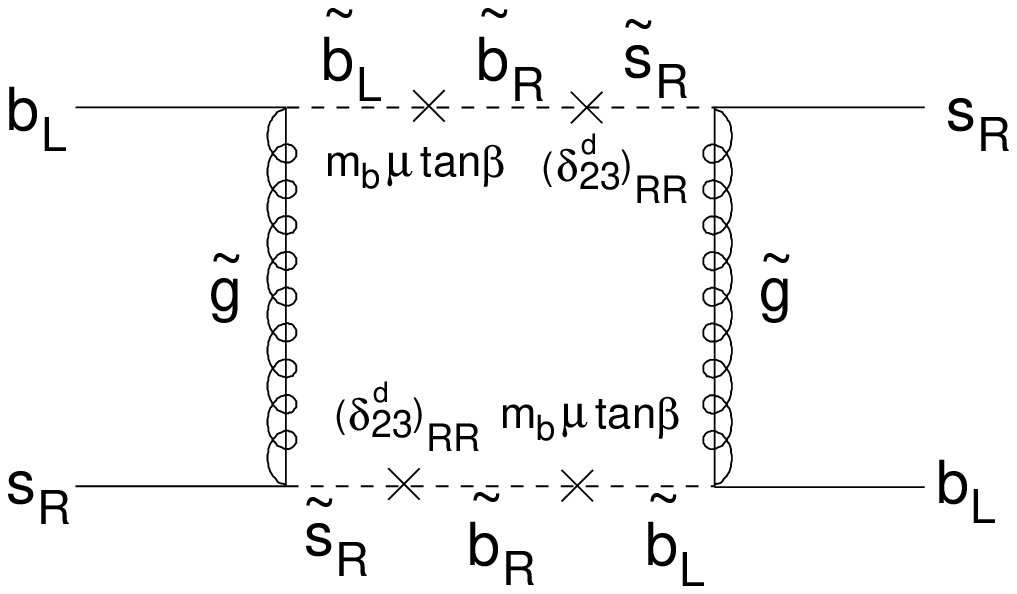}
  \caption{Diagrams contributing to $B_s$-mixing through large
    $\tilde{b}_R$-$\tilde{s}_R$ mixing. The diagram on the right induces 
    effective $\tilde{b}_{L}$-$\tilde{s}_{R}$ mixing which contributes to
    $C_{LL}^{S}$ and $C_{LL}^{S\times}$.}
\label{fig:diagBmix}
\end{figure}

Mixing between $\tilde{b}_R$ and $\tilde{s}_R$ also leads to a significant
contribution to $B_s$-$\bar{B}_s$ mixing. In our scenario the effective
Hamiltonian that receives such contributions consists of three
operators that have nonzero coefficients:
\begin{equation}
\label{BmixHeff}
\mathcal{H}_\mathrm{eff}^{SUSY}=C^V_{RR}(\sbar\gamma^\mu P_R b)
(\sbar\gamma_\mu P_R b) + C^S_{LL}(\sbar P_L b)(\sbar P_L b) +
C^{S\times}_{LL}(\sbar_i P_L b_j)(\sbar_j P_L b_i).
\end{equation}
The Wilson coefficients at the high scale are obtained
by matching the effective Hamiltonian to the $\Delta B=2$ squark-gluino box
diagrams like those shown in Figure~\ref{fig:diagBmix}. 
The result is given by\footnote{This result differs from 
\cite{Hagelin:1992tc}, but agrees with subsequent analyses, 
e.g. \cite{Gabbiani:1996hi}.}:
\begin{eqnarray}
\label{eqn:BmixWilson1}
C^V_{RR}&=&\frac{\alpha_s^2}{m_{\tilde{g}}^2}\sum_{AB}\Gamma^{R*}_{sA}
\Gamma^{R}_{bA}\Gamma^{R*}_{sB}\Gamma^{R}_{bB}
\times \left[ -\frac{11}{9}B_1(x_A,x_B) -\frac{1}{9}B_2(x_A,x_B)
\right] \\
\label{eqn:BmixWilson2}
C^S_{LL}&=&\frac{\alpha_s^2}{m_{\tilde{g}}^2}\sum_{AB}\Gamma^{R*}_{sA}
\Gamma^{L}_{bA}\Gamma^{R*}_{sB}\Gamma^{L}_{bB}
\times \left[ -\frac{17}{18}B_2(x_A,x_B) \right] \\
\label{eqn:BmixWilson3}
C^{S\times}_{LL}&=&\frac{\alpha_s^2}{m_{\tilde{g}}^2}\sum_{AB}\Gamma^{R*}_{sA}
\Gamma^{L}_{bA}\Gamma^{R*}_{sB}\Gamma^{L}_{bB}
\times \left[ \frac{1}{6}B_2(x_A,x_B) \right],
\end{eqnarray}
where the loop-functions are defined 
in Appendix~\ref{loopfn}, the $\Gamma$'s are defined in 
Equation~(\ref{eqn:gammadef}), and again $x_A = \tm_{A}^2/m^2_{\tilde{g}}$.

The leading Standard Model contribution to $B_s$ mixing is induced by
a top quark box diagram which yields the following effective
Hamiltonian \cite{Buchalla:1995vs}
\begin{equation}
\mathcal{H}_\mathrm{eff}^{SM}=C_{LL}^{SM}(\sbar\gamma^\mu P_L b)
(\sbar\gamma_\mu P_L b),
\end{equation}
with the Wilson coefficient matched at $M_W$
\begin{equation}
C_{LL}^{SM}=\frac{G_F^2}{16\pi^2}M_W^2 
\left(V_{ts}^*V_{tb}\right)^2 S(m_t^2/M_W^2),
\end{equation}
where
\begin{equation}
S(x)=\frac{4x-11x^2+x^3}{4(1-x)^2}-\frac{2x^3\ln x}{2(1-x)^3}.
\end{equation}

Before taking the hadronic matrix element of the effective
Hamiltonian we must first take QCD corrections into account by using
the renormalization group equations (RGE) to evolve the Wilson
coefficients down to the low scale. The general NLO
running of the Wilson coefficients for a $\Delta B=2$ effective
Hamiltonian is given in \cite{Buras:1990fn} and \cite{Becirevic:2001jj} and
involves mixing among different coefficients. In our case only the 
two scalar left-left operators mix, while the vector right-right and the  
vector-left-left coefficients simply scale multiplicatively. 
For simplicity we have evolved both operators from $M_W$ down to 
the mass of the $b$-quark.\footnote{The supersymmetric contribution
should in fact run from 
$\sim m_{\tilde{g}}$ down to $m_b$. In this approximation we are ignoring 
corrections of order $1-\alpha_s(m_{\tilde{g}})/\alpha_s(M_W)$ and potential 
contributions from the top quark in loops which is smaller. These corrections 
are part of the systematic uncertainty in our calculation.}

The hadronic matrix element of the effective Hamiltonian between $B_s$
and $\bar{B}_s$ states was calculated on the lattice \cite{Becirevic:2001xt}
\bea
\langle \bar{B}_s \left|(\sbar\gamma^\mu P_R
b)(\sbar\gamma_\mu P_R b)\right|B_s\rangle &=&
\frac{1}{3}m_{B_s}f^2_{B_s}B_1(\mu) \label{eqn:matrixRR} \\
\langle \bar{B}_s \left|(\sbar P_L b)(\sbar P_L b)\right|B_s\rangle &=&
-\frac{5}{24} \left( \frac{M_{B_s}}{m_b+m_s}\right)^2
m_{B_s}f^2_{B_s}B_2(\mu) \\
\langle \bar{B}_s \left|(\sbar_i P_L b_j)(\sbar_j P_L b_i)\right|B_s\rangle &=&
\frac{1}{24} \left( \frac{M_{B_s}}{m_b+m_s}\right)^2
m_{B_s}f^2_{B_s}B_3(\mu), 
\eea
with $f_{B_{s}}$ as given in  
Table~\ref{tab:inputBmix}, $B_1(m_b)=0.87(4)^{+5}_{-4}$,
$B_2(m_b)=0.82(3)(4)$, and
$B_3(m_b)=1.02(6)(9)$, where the first error is statistical and the
second is systematic, excluding uncertainty due to quenching.  The quark 
masses in the above expression should be evaluated at the scale $\mu$. 
The hadronic matrix element for
the left-left current operator of the Standard Model is identical to that of 
the right-right operator shown in Equation~(\ref{eqn:matrixRR}).
Finally, we can write the expression for the mass difference between
$B_s$ and $\bar{B}_s$ as
\begin{equation}
\label{deltams}
\Delta M_{B_s}=2\left|\langle
\bar{B}_s\left|\mathcal{H}_\mathrm{eff}^{\Delta B=2}
\right|B_s \rangle \right|.
\end{equation}
The Standard Model and supersymmetric contributions interfere,
$\mathcal{H}_\mathrm{eff}^{\Delta 
B=2}=\mathcal{H}_\mathrm{eff}^{SM}+\mathcal{H}_\mathrm{eff}^{SUSY}$.

The input parameters used in the calculation are given in Table
\ref{tab:inputBmix}. Our results should be compared to the standard
model prediction which can be obtained roughly by taking 
$\mathcal{H}_\mathrm{eff}^{\Delta B=2}=\mathcal{H}_\mathrm{eff}^{SM}$ 
in Equation~(\ref{deltams}), which yields $\Delta M_{B_s}^{SM}\sim
13.8\mbox{ ps}^{-1}$. 
A more rigorous treatment given in \cite{Ciuchini:2000de} yields 
\begin{equation}
\label{SMpredict}
\Delta M_{B_s}^{SM}=17.3^{+1.5}_{-0.7} \mbox{ ps}^{-1}.
\end{equation}
However, given the substantial uncertainty in the lattice evaluation of,
e.g., $f_{B_{s}}$, it is probably appropriate to inflate this error, likely to 
the 25$\%$ level \cite{Kronfeld:2002ab}.
The current experimental limit, combining results from the LEP experiments 
and SLD, is \cite{LEPBs}
\begin{equation}
\Delta  M_{B_s}>14.4 \mbox{ ps}^{-1} \mbox{ (95 \% confidence level)}.
\end{equation}
Current and upcoming experiments are expected to be sensitive to mass
differences much greater than the  
Standard Model prediction shown in Equation~(\ref{SMpredict}). 
At Run II of the Tevatron \cite{Anikeev:2001rk} CDF is expected to probe
 up to $\Delta M_{B_s}$ of 41 ps$^{-1}$ while BTeV is expected to achieve
sensitivity to values up to $\Delta M_{B_s} \sim 55$ ps$^{-1}$. 
Any evidence that $\Delta M_{B_s} > 25$ ps$^{-1}$ from these experiments
would be a clear signal of new physics.

\begin{table}[tbp]
  \begin{center}
    \begin{tabular}{|c|c||c|c||c|c|}
      \hline 

      Parameter & Value & Parameter & Value & Parameter & Value\\
      \hline \hline
      $m_b$ &4.2 GeV & $m_{B_s}$ & 5.379 GeV & $V_{ts}^*V_{tb}$ & $-$0.04\\
      $m_t$ & 174 GeV & $f_{B_s}$ & 204 MeV & $\alpha_s(M_Z)$ & 0.1185  \\
      $M_W$ & 80.4 GeV & $M_Z$ & 91.2 GeV& $\tau_{B_s}$ &  1.461 ps  \\

      \hline
    \end{tabular}
    \caption{Input parameters used in the calculation of $S_{\phi K}$
and $\Delta M_{B_s}$.}
    \label{tab:inputBmix}
  \end{center}
\end{table}

To illustrate our results we add contours of constant
$\Delta M_{B_s}$ (red dashed lines)
to Figures~\ref{fig:contour1}-\ref{fig:lrcontour1}.  
In the case of dominant right-right contributions, i.e. small $\mu
\tan \beta$, 
the trend is similar to that of the previous section; lighter gluino and 
squarks give a larger SUSY contribution and thus increase $\Delta M_{B_s}$.
Note that the supersymmetric contribution to the mass difference 
dominates over the Standard Model in significant regions of the 
supersymmetric parameter space, easily allowing 
$\Delta M_{B_s}>100 \mbox{ ps}^{-1}$ where right-right mixing
dominates. Such values for $\Delta M_{B_s}$ are
certainly beyond the reach of the experiments mentioned above. 

In the case of dominant $LR+RR$ mixing the modification of $B_{s}$ mixing 
is not as striking. In the example given in Figure~\ref{fig:lrcontour1} values 
of $\Delta M_{B_s}$ are much closer to the standard model
prediction. Restricting ourselves to areas that respect the $b \to s
\gamma$ bound gives a yet lower 
value, within the reach of upcoming experiments. We should point out that 
this is not generic since Figure~\ref{fig:lrcontour1} only represents a slice 
of parameter space. 
Other choices of parameters can give higher values of $\Delta M_{B_s}$ 
(above 30 $\mbox{ps}^{-1}$) for high values of $\mu \tan \beta$. The 
correlation of these results to those in $B_{d}^{0} \to \phi K_S$ will be 
discussed in the next section.  

Finally, we should comment about possible CP violation in the 
$B_s$-$\bar{B}_s$ system. In the standard CKM scenario the  
$B_s \rightarrow \bar{B}_s$ amplitude does not have a CP violating phase 
(in the Wolfenstein parameterization), so no indirect CP violation is expected.
In our scenario, however,
$B_s$ mixing can involve the phases from the down-squark mass matrix. In the 
cases where the SUSY contribution to $B_s$ mixing dominates the SM,
measurements of CP violation in  $B_s \to J/\psi \phi$ 
will be sensitive to these phases.

%%%%%%%%%%%%%%%%%%%%%%%%%%%%%%%%%%%%%%%%%%%%%%%%%%%%%%%%%%%%%%%%%%%

\section{Correlation}
In this section we will discuss the correlation between 
$S_{\phi K}$ and $\Delta M_{B_{s}}$ in the context of large
$\tilde{b}$-$\tilde{s}$ mixing.
Because the effect on $B_s$ mixing is
very different for the two limiting regions of parameter space, we
will discuss them separately. For related studies see
\cite{Lunghi:2001af,Grossman:1999av}. 

\subsection{Dominant $RR$ Mixing}  
In this region of parameter space the operators $\mathcal{O}'_{3-6}$
make large contributions to $S_{\phi K}$, while there is essentially
only one contribution to 
$B_{s}$ mixing, namely that from the operator shown in 
Equation~(\ref{eqn:BmixWilson1}). Unfortunately there is 
no simple, precise, relationship between the combination of the
$\Delta B=1$ 
operators and the operator responsible for $B_{s}$ mixing.  
In general, they depend quite differently on loop functions.

In spite of this, one can make the following strong statement. In cases where 
there is a large shift in $S_{\phi K}$ away from the Standard 
Model expectation due to the operators $\mathcal{O}'_{3-6}$, and the
$RR$ contribution to $\mathcal{O}'_{g}$ (the
dominant right-right mixing scenario), there is
a large contribution to $B_{s}$ mixing.  
To see this, we first note that the squarks and gluino must 
not be too heavy, and the $\tilde{b}$-$\tilde{s}$ mixing must be large in 
order to have a large contribution to 
$S_{\phi K}$.  This suggests a minimum contribution to
the $B_{s}$ mixing.  However, there is the worry that it might
be possible to fine-tune parameters to somehow drastically suppress the 
contribution to $B_{s}$ mixing; for example, by choosing squark and gluino 
mass ratios to minimize the value of the functions $B_1$ and $B_2$ in 
Equation~(\ref{eqn:BmixWilson1}).  We find that this is not possible, 
however.  In order to have a very large contribution to $S_{\phi K}$, 
one is pushed into a region of parameter
space where the gluino, and at least one of the down-type squarks is light. 
Furthermore, the splitting between this light squark, which represents
a mixture of 
$\tilde{b}$ and $\tilde{s}$ squarks, and the masses of the heavier 
squarks must be 
large to avoid a super-GIM cancellation.  Once this qualitative picture 
for the spectrum 
is identified, it is easy to check that there cannot be a cancellation 
of the contribution to $B_{s}$ mixing in this case.

Now we present our results quantitatively.  Since our goal here will be to 
show that large deviations in $S_{\phi K}$ will correspond to large
contributions to $B_{s}$ mixing, we plot the minimum achievable value 
of $\Delta M_{B_s}$ for a given value of $S_{\phi K}$.  The minimum is found 
by scanning a parameter space that consists of the parameters 
$\{ \mgluino, \tm_{L2} , \tm_{R2}, \tm_{R3}, \cos \theta_{56}, \phi_6 \}$.    
As discussed in Equation~(\ref{eqn:gammadef}), $\cos \theta_{56}$
represents the
mixing angle between the right-handed $\tilde{b}$ and $\tilde{s}$ squarks, and 
$\phi_6$ represents the phase corresponding to this off-diagonal term.
As a parameter space, we take:\footnote{Here we have taken
$\mu\tan\beta=0$ for simplicity. Nonzero values can weaken the
correlation somewhat, as shown in Figure~\ref{fig:maxminmbs}.}
\begin{eqnarray}
\label{eqn:params}
\mgluino &\in& (200,700) \mbox{ GeV} \\
\tm_{L2}&\in& (300,2500)  \mbox{ GeV} \nonumber \\ 
\tm_{R2}& \in& (300,2500) \mbox{ GeV}\nonumber  \\ 
\tm_{R3}& \in& (250,1000) \mbox{ GeV}\nonumber  \\ 
\theta_{56} &\in& (0, \pi/2) \nonumber\\ 
\phi_6 &\in& (0, 2\pi). \nonumber 
\end{eqnarray}
The lower limits on the masses are motivated by direct searches, 
while the upper limits are motivated by naturalness considerations.
A scan would generate a scatter plot of 
$S_{\phi K}$ vs. $\Delta M_{B_s}$.  For a given resultant value 
of $S_{\phi K}$, we find the combination 
of parameters that yields the smallest contribution to $B_{s}$ mixing.
This is essentially equivalent to taking the boundary of the region 
generated by the scatter plot.  

\begin{figure}
\centering
\includegraphics[width=.6\textwidth]{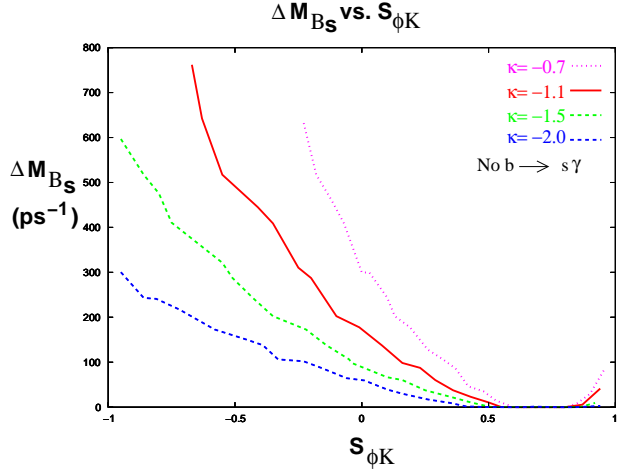}
\caption{The minimum value of $\Delta M_{B_s}$ for a given value 
of $S_{\phi K}$.  It is found by scanning over the parameter space given in
Equation~(\ref{eqn:params}).  The corresponding curve is shown for 
several values of $\kappa$, the coefficient of the dipole operator, as defined
in Equation~(\ref{eqn:kappadef}).}
\label{fig:nobsg}
\end{figure}

\begin{figure}
\centering
\includegraphics[width=.6\textwidth]{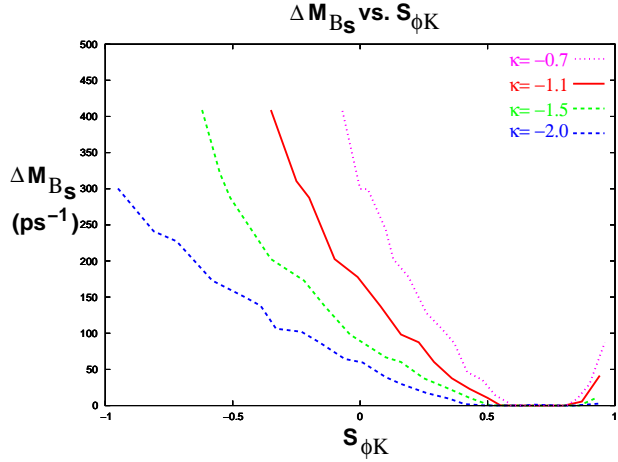}
\caption{The same as Figure~\ref{fig:nobsg}, but with the 
$b \rightarrow s \gamma$ constraint applied as discussed in the text.  
The minimum value of $\Delta M_{B_s}$ for a given value 
of $S_{\phi K}$.  It is found by scanning over the parameter space given in
Equation~(\ref{eqn:params}).  The corresponding curve is shown for 
several values of $\kappa$, the coefficient of the dipole operator, as 
defined in Equation~(\ref{eqn:kappadef}).}
\label{fig:bsg}
\end{figure}

As discussed in Section \ref{sec:bstrans}, there is considerable 
dependence on the variable $\kappa$, which has a relatively large
 uncertainty.  So we repeat the above exercise for several values 
of $\kappa$,
displaying the results in Figure~\ref{fig:nobsg}. Adding the
constraint from $b \rightarrow s \gamma$ modifies these results as shown in
Figure~\ref{fig:bsg}.  The contours in 
Figure~\ref{fig:bsg} notably do not extend as low in 
$S_{\phi K}$ because the $b \rightarrow s \gamma$ constraint removes the 
region of parameter space that allowed us to obtain those values in 
Figure~\ref{fig:nobsg}.

The take-home message from the figures is a simple one.  If the hint 
of  the deviation in $S_{\phi K}$ measured in the 
$B \rightarrow \phi K_{S}$ persists and it is  attributable to 
a scenario with dominant $RR$ squark mixing, it will result in a 
large contribution to $B_{s}$ mixing, which will be a clear 
indication of new physics observable at the Tevatron.

\subsection{Dominant $LR+RR$ Mixing}

In the region of parameter space where $\mu\tan\beta$ is relatively
large, the expectation for $\Delta M_{B_s}$ is very different. In this
region the main contribution to $S_{\phi K}$ comes from the $LR+RR$
contribution to the dipole operator $\mathcal{O}'_{g}$. 
This contribution can be sizeable even when the squarks and gluinos are
heavy (squarks can be at the TeV level or higher). 
This is the significant difference between the two limiting cases.
Heavy squarks and gluino mean
that the contributions from the operators $\mathcal{O}'_{3-6}$ are
small. Similarly, the operators responsible for $B_s$ mixing, which
come from box diagrams and resemble $\mathcal{O}'_{3-6}$, can also be 
small. The bottom line is that a large contribution to $S_{\phi K}$ is 
possible without a large addition to $\Delta M_{B_s}$. 
This is borne out numerically, as shown in Figure~\ref{fig:lrcontour1},
where the  points allowed by the $b\rightarrow s\gamma$ constraint all
give $\Delta M_{B_s}$ very close to the Standard Model
expectation.

\begin{figure}[tbp]
  \centering
  \includegraphics[width=0.49\textwidth]{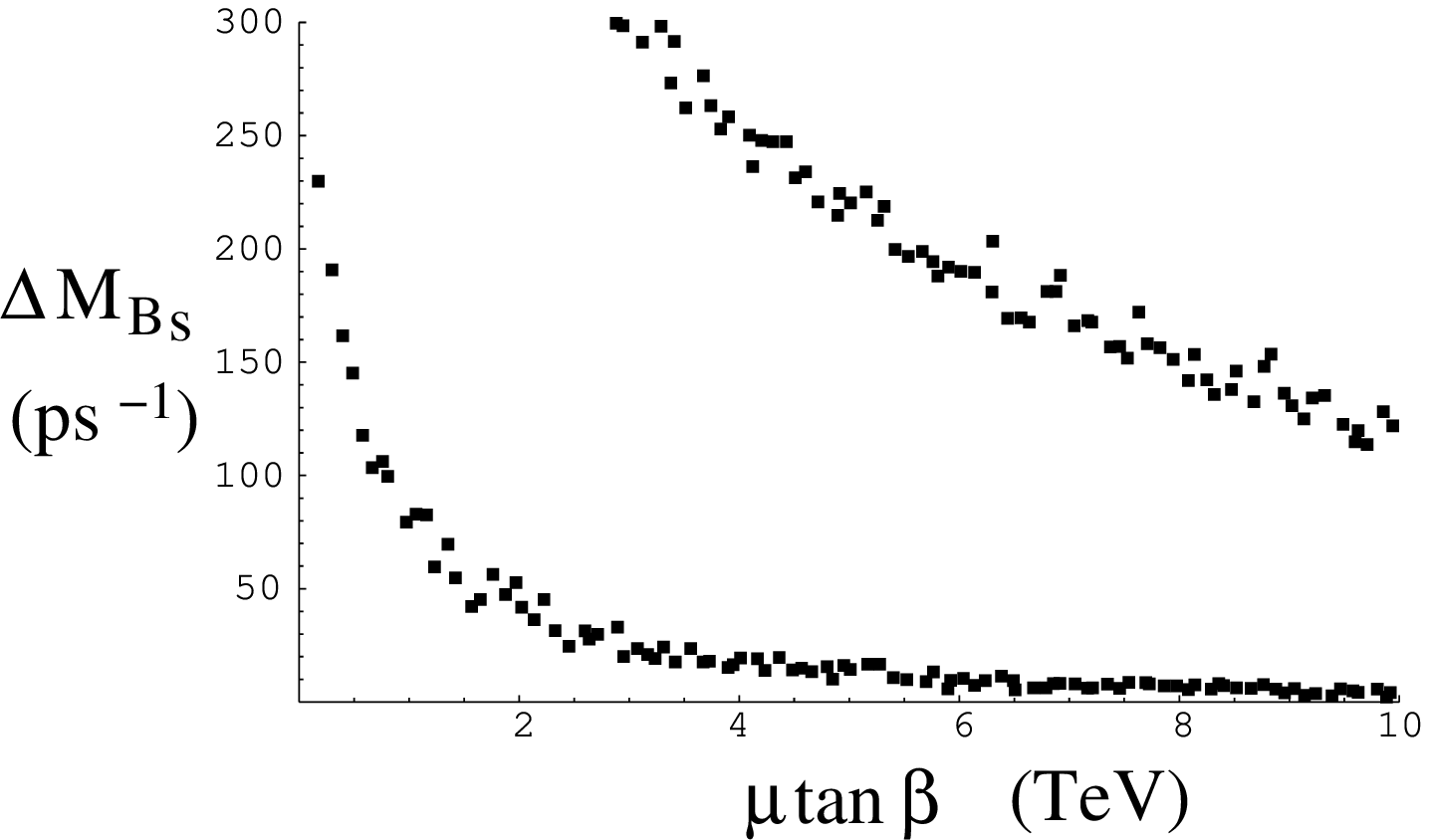}
  \includegraphics[width=0.49\textwidth]{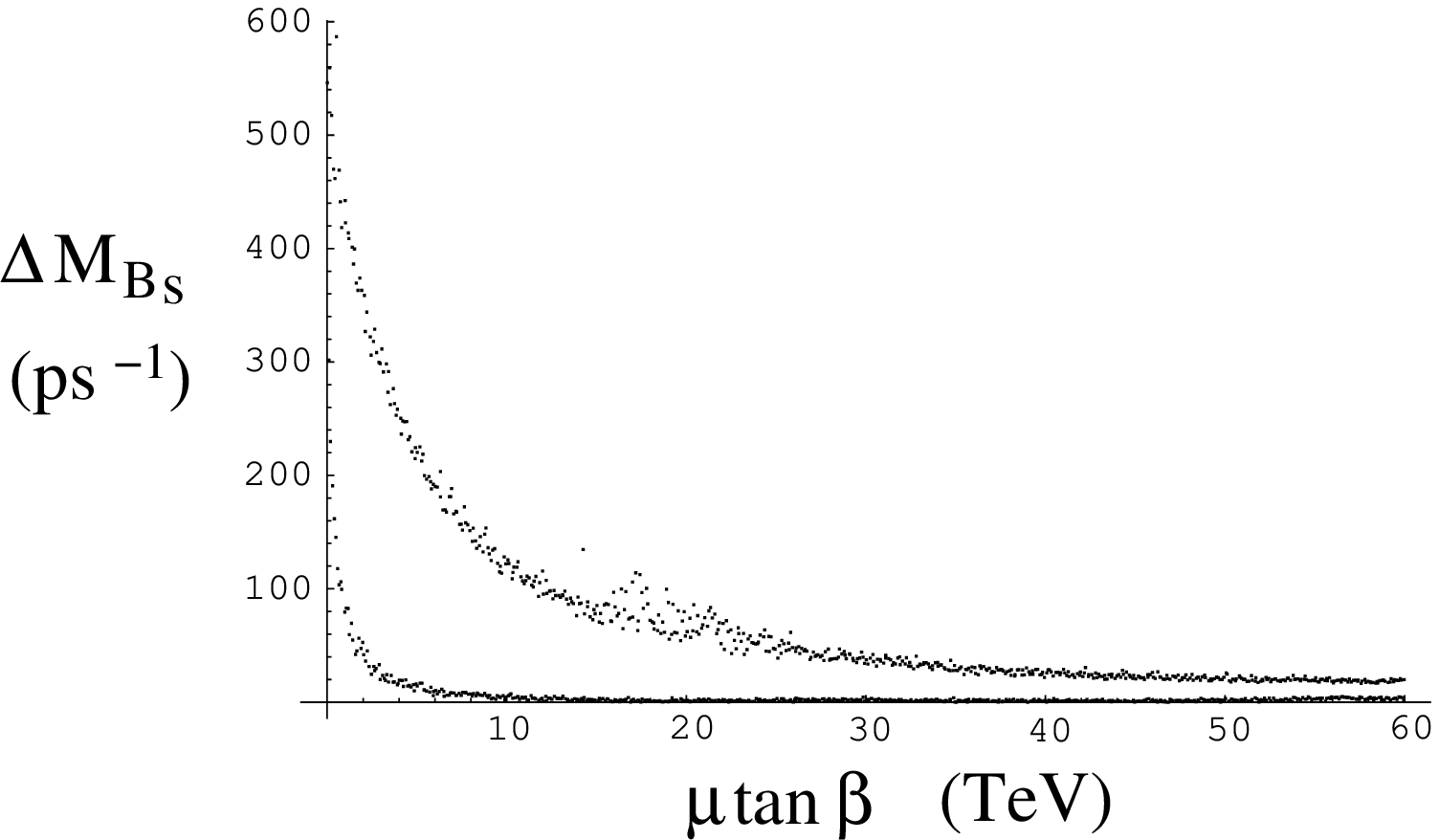}
  \caption{Maximum and minimum values of $\Delta
M_{B_s}$ as a function of $\mu\tan\beta$ resulting from a scan of
150000 points in the parameter space with $S_{\phi K}<-0.2$
and $\kappa = -1.1$. The plot on the left shows an enlargement of the
region of small $\mu\tan\beta$.} 
\label{fig:maxminmbs}
\end{figure}

We have seen that the $LR+RR$ contribution to $S_{\phi K}$ can be important
even for fairly small values of $\mu\tan\beta$, so it is natural to
wonder what conclusions can be drawn about $\Delta M_{B_s}$ in the
regions where both contributions are important. To answer this
question we again performed a scan of the parameter space, this time
collecting points with maximal and minimal $\Delta M_{B_s}$ as a
function of $\mu\tan\beta$ with the additional requirement that
$S_{\phi K} < -0.2$ for the nominal value $\kappa = -1.1$. The results
are shown in Figure~\ref{fig:maxminmbs}. In accord with what was
stated above, points with the largest $\mu\tan\beta$ give smaller
contributions to $B_s$ mixing. In fact, for $\mu\tan\beta > 40$ TeV any
effects on $\Delta M_{B_s}$ will be indistinguishable from the Standard
Model expectation. This apparent upper bound may be interpreted as follows.
For large $\mu \tan \beta$ the severe $b\to s \gamma$ bound is pushing
us to regions where $RR$ mixing is small or masses are high, both of
which disfavor large contributions to $B_s$ mixing. 
At the other end of the
spectrum, for $\mu\tan\beta < 2.4$ TeV all points in the scan gave values
of $\Delta M_{B_s} > 30$ ps$^{-1}$, a clear signal of new physics
above even the largest Standard Model predictions. This trend
continuously connects us back to the result of the previous
subsection.
%\footnote{For larger values of $\kappa$ the correlation
%with $B_s$ mixing is weaker, but even for $\kappa=-2.0$ and $S_{\phi
%K}<-0.9$ we find $\Delta M_{B_s} >30$ ps$^{-1}$ for all points with
%$\mu\tan\beta < 1.5$ TeV.}

Mixing between the first two generations
of squarks has fallen outside the main scope of this paper.  
We should mention, however, that larger values of the lightest squark 
masses, around 500 GeV, may well be preferred by constraints from 
$K$-$\bar{K}$ mixing.  To see this, note that 
if we believe that the Cabibbo angle originates through the down Yukawa 
matrices, then the down Yukawa matrix has the structure:
\begin{equation}
\mathbf{\lambda}_{d}=\left( 
\begin{array}{cc} 
h_{d} & h_{s} \lambda_{C} \\
h & h_{s}  \\
\end{array} \right),
\end{equation}
where $h$ is some unknown Yukawa coupling and $\lambda_{C}$ is the
sine of the Cabibbo angle.
Diagonalizing this matrix requires a rotation on $d_{L}$ and $s_{L}$
of ${\mathcal O}(\lambda_{C})$ which induces an off-diagonal element
of order $h_{d} \lambda_{C}$.
Using the phenomenological relationship, $m_{d}/m_{s} \approx \lambda_{C}^{2}$,
we find that a rotation between $d_{R}$ and $s_{R}$ 
of ${\mathcal O}(\lambda_{C}^{3})$ is needed to complete the diagonalization.
  Then, due to the lack 
of degeneracy between the squarks, the induced $\tilde{d}$-$\tilde{s}$ mixing 
can lead to a large contribution 
to the $K$-$\bar{K}$ mixing.  This suggests that heavier squark masses, perhaps
above 500 GeV are preferred, barring some accidental cancellation with the
(unknown) $(1,2)$ element, $h$, in the above matrix. 

By the above reasoning, if one wishes to achieve an $S_{\phi K}$ that
differs significantly from the value of $\sin 2 \beta$ as measured in
the in the 
$B_{d}^{0}\rightarrow J/\psi K_{S}$ channel, there may a  
theoretical prejudice to prefer scenarios where
% where the $LR+RR$ 
%contributions dominate $S_{\phi K}$, since in that case
large squark masses are more easily accommodated, such as the $LR+RR$
dominant case or the $RR$ dominant case in conjunction with a large
value of $\kappa$.
% while still giving a 
%large effect due to the $\mgluino/m_{b}$ enhancement.

\section{Conclusion}
There exist a class of models, motivated by Grand Unified theories and 
the large observed mixing in atmospheric neutrinos, where it is natural 
to have a large mixing between the right-handed $\tilde{b}$ squark 
and $\tilde{s}$ squark.  
We have found that there exists a range of parameters where such mixing 
induces a significant deviation in $S_{\phi K}$
from the Standard Model expectation of $\sin 2\beta$ as measured in the
channel $B_{d}^{0} \rightarrow J/\psi K_{S}$.  

In particular, the central value for $S_{\phi K}$ from BaBar and BELLE can 
be accommodated 
without conflicting with the measured value of $b \rightarrow s
\gamma$, using the naive estimate for $\kappa$, the hadronic matrix 
element for the chromo-dipole operator. For larger values of $\kappa$
any value of $S_{\phi K}$ is allowed within the constraint.

There are two possible origins of a substantial modification to $S_{\phi K}$.
The first solely involves a large right-right mixing, with no 
contribution from the mixing proportional to $m_{b} \mu \tan \beta$.  
In this case a 
small gluino mass is required, near the experimental bound.  
Correspondingly, there is a large contribution to $B_{s}$ mixing, a
consequence which will be
testable at the Tevatron Run II.  In the second case, we 
consider the mixing from the combination of 
the large right-right mixing and a large $m_{b} \mu \tan \beta$.  
In this case, squarks and gluinos need not be light, so 
$B_{s}$ mixing need not be large.  
In particular, for very large values of $\mu \tan \beta$ the prediction 
for $B_{s}$ mixing is indistinguishable from the Standard Model 
prediction, when current errors on lattice matrix elements are taken into 
account.  However, a substantial improvement in the Standard Model
prediction for $B_{s}$ mixing still may allow an effect to be seen at the 
Tevatron in this case.

\section*{Note Added}
While completing this paper we received References
\cite{Kane:2002sp,Khalil:2002fm}. There is some 
overlap with these papers, which also consider supersymmetric contributions
to $B_{d}^{0} \to \phi K_{S}$. 

Regarding \cite{Kane:2002sp}, in places that we overlap, we agree
qualitatively 
with their results, though there may be some quantitative differences.  
These are likely due to the fact that they work in the mass insertion 
approximation. Indeed, allowing large mixings and hierarchies that cannot 
be described by mass insertions gives us larger contributions in 
the `pure' $RR$ mixing case.  Other possible differences may arise from 
a different treatment of the hadronic matrix elements and the fact that 
constraints from $b \to s \gamma$ were not imposed in the same way.

%We appear to disagree with
%\cite{Khalil:2002fm}, which seems to find a strong
%suppression of the contribution from the $\delta_{23}^{RR}$ mass
%insertion.\footnote{This result was subsequently modified.}

We differ from both papers in our emphasis on the mixing induced by a
combination of $m_{b} \mu \tan \beta$ along with a large
flavor-changing $RR$ element in the squark mass matrix. Though double
mass insertions are briefly discussed in \cite{Kane:2002sp}, both
\cite{Kane:2002sp} and \cite{Khalil:2002fm} focus on the contributions
of single mass insertions, including flavor off-diagonal $LR$ mixing.

We would like to emphasize that the double mass insertion ($LR+RR$
mixing) does not 
necessarily describe the same physics as a single $LR$ flavor mixing
insertion, a point also mentioned in \cite{Kane:2002sp}. Treating a $LR+RR$
mixing as a pure $\delta_{23}^{RL}$ may miss important contributions
due to the $RR$ mixing only. For example, we find maximal values of
$\Delta M_{B_s}$ for intermediate values of $\mu \tan \beta$ which are
much higher than those plotted in
\cite{Kane:2002sp} with pure $LR$ mixing. This might be due to a $RR$ mixing
contribution which is sub-dominant in $S_{\phi K}$ but which
nevertheless gives a large contribution to $\Delta M_{B_s}$.  This
example illustrates how our framework differs from analyses which
consider only one mass insertion at a time.

\section*{Acknowledgments}
We would like to thank Zoltan Ligeti, Gustavo Burdman, Yuval Grossman,
and Gudrun Hiller for useful conversations.

This work was supported in part by the Director, Office of Science,
Office of High Energy and Nuclear Physics, Division of High Energy
Physics of the U.S. Department of Energy under Contracts
DE-AC03-76SF00098 and DE-AC03-76SF00515 and in part by the National 
Science Foundation under grant PHY-0098840.

\appendix

\section{Loop Functions}
\label{loopfn}

We include the loop functions for completeness. We use the same
definitions as in~\cite{Moroi:2000tk}. Here $x_A\equiv
m_{\tilde{d}_A}^2/m_{\tilde{g}}^2$.  
\begin{eqnarray}
    B_1(x_A,x_B) &=&  \int_0^\infty \frac{-\frac{1}{4} z^2 \, dz}{(z+1)^2
                (z+x_A) (z+x_B)} \nonumber \\ 
        &=& 
    -\frac{x_A^2 \log x_A}{4(x_A-x_B)(x_A-1)^2}
    -\frac{x_B^2 \log x_B}{4(x_B-x_A)(x_B-1)^2}
    \nonumber \\ &&
    -\frac{1}{4(x_A-1)(x_B-1)},
    \\
    B_2(x_A,x_B) &=& \int_0^\infty \frac{z \, dz}{(z+1)^2
                (z+x_A) (z+x_B)} \nonumber \\ 
        &=& 
    -\frac{x_A \log x_A}{(x_A-x_B)(x_A-1)^2}
    -\frac{x_B \log x_B}{(x_B-x_A)(x_B-1)^2}
    \nonumber \\ &&
    -\frac{1}{(x_A-1)(x_B-1)},
\end{eqnarray}
\begin{eqnarray}
    C_1 (x) &=&
    \frac{2x^3-9x^2+18x-11-6\log x}{36(1-x)^4},
    \\
    C_2 (x) &=&
    \frac{-16x^3+45x^2-36x+7+6x^2(2x-3)\log x}{36(1-x)^4},
\end{eqnarray}
\begin{eqnarray}
    D_1 (x) &=&
    \frac{-x^3+6x^2-3x-2-6x\log x}{6(1-x)^4},
    \\
    D_2 (x) &=&
    \frac{-x^2+1+2x\log x}{(x-1)^3},
    \\
    D_3 (x) &=&
    \frac{2x^3+3x^2-6x+1-6x^2\log x}{6(1-x)^4},
    \\
    D_4 (x) &=&
    \frac{-3x^2+4x-1+2x^2\log x}{(x-1)^3}.
\end{eqnarray}

\section{Chromo-dipole Matrix Element for $B_{d}^{0} \rightarrow \phi K_S$}
\label{matrix-element}
In this appendix we show explicitly the computation of the matrix
element of $\mathcal{O}^{(\prime)}_g$ in the naive factorization
approximation. The computation of a similar quantity for $\bar{B}^0
\rightarrow \pi^+ \pi^-$ decay can be found in~\cite{Arhrib:2001gr}.

We start with
\begin{equation}
\mathcal{O}_{g} = \frac{g_s}{8\pi^2} m_b (\sbar_i \su T^a_{ij}
P_R b_j)G^a_{\mu\nu} 
\end{equation}
and then connect a quark current through a virtual gluon to form a
four-quark operator.  This step depends on the
convention used for the covariant derivative.\footnote{In particular,
note that~\cite{Deshpande:1995tx} appears to use the opposite convention
of~\cite{Arhrib:2001gr,Kagan:1997qn}.} Our convention is that $D_\mu = \partial_\mu
+ igT^a A^a_\mu$, and we have checked that this is consistent with the
Wilson coefficients for both the standard model and SUSY
contributions. This yields the operator
\begin{equation}
\mathcal{O}_{g4} =
i\frac{\als}{\pi} \frac{k_\nu}{k^2} m_b (\sbar_i \su T^a_{ij} P_R b_j)
(\bar{q}_k \gd T^a_{k\ell} q_\ell)
\end{equation}
where $k=p_b-p_s$ is the gluon momentum.
In the naive factorization 
approximation the color-octet current, $(\bar{q}_k \gd T^a_{k\ell} q_\ell)$, 
cannot produce a physical $\phi$, so the $\phi$ must be produced by the 
$\sbar$ and $q$ operators. To factor the matrix element we first use the 
equations of motion to simplify the tensor current. This yields:
\begin{eqnarray}
\mathcal{O}_{g4}= -\frac{\als}{\pi k^2} m_b \left\{ m_b(\sbar\gu P_L
T^a_{ij} b) + 
m_s(\sbar\gu P_R T^a_{ij} b) - 2p_b^\mu(\sbar_i P_R T^a_{ij} b_j)
\right\} \times \nonumber \\
(\bar{q}_k \gd T^a_{k\ell} q_\ell)
\end{eqnarray}
where the last term was simplified using the conservation of the quark
current, $k^\mu (\bar{q}_k \gd T^a_{k\ell} q_\ell)=0$ as
in~\cite{Arhrib:2001gr}. Then by a Fierz transformation and judicious use of Dirac
matrix identities, this can be brought to the
form~\cite{Barbieri:1997kq,Deshpande:1995tx,Kagan:1997qn} 
\begin{eqnarray}
\mathcal{O}_{g4} &=& -\frac{\als}{2\pi k^2} m_b \frac{N_c^2-1}{2N_c^2}
\left( \delta_{i\ell}\delta_{kj}-\frac{2N_c}{N_c^2-1} T^a_{i\ell}
T^a_{kj} \right)  \\
& & \times\left[ 2m_b(\sbar_i\gu P_L q_\ell)(\bar{q}_k \gd P_L b_j)
-4m_b(\sbar_i P_R q_\ell)(\bar{q}_k P_L b_j) \right. \nonumber \\
& & \ \ \ \ +2m_s(\sbar_i\gu
P_R q_\ell)(\bar{q}_k \gd P_R b_j) -4m_s(\sbar_i P_L
q_\ell)(\bar{q}_k P_R b_j) \nonumber \\
& &\ \ \ \ +2(p_b)_\mu \left\{ (\sbar_i\gu P_L q_\ell)(\bar{q}_k
P_R b_j) + (\sbar_i P_R q_\ell)(\bar{q}_k \gu P_R b_j) \right. \nonumber \\
& & \ \ \ \ \ \ \ \ \left. \left. +i(\sbar_i\su
P_R q_\ell)(\bar{q}_k \gamma^\nu P_R b_j) -i (\sbar_i\gamma_\nu P_L
q_\ell)(\bar{q}_k \su P_R b_j) \right\} \right] \nonumber  
\end{eqnarray}
Next we use the following parameterization for matrix elements.
\begin{eqnarray}
\langle \phi(p_\phi,\epsilon_\phi)|\sbar \gu s|0\rangle &=& f_\phi m^2_\phi
\epsilon^\mu_\phi \\
\langle \phi(p_\phi,\epsilon_\phi)|\sbar s|0\rangle &=& 0 \\
\label{tensor} \langle \phi(p_\phi,\epsilon_\phi)|\sbar \su s|0\rangle
&=& -if_\phi m^2_\phi 
\frac{2m_s}{m_\phi^2} (\epsilon^\mu_\phi p^\nu_\phi -
\epsilon^\nu_\phi p^\mu_\phi).
\end{eqnarray}
Also,
\begin{eqnarray}
 \langle \bar{K}^0(p_K)|\sbar \gu b|\bar{B}^0_d(p_B)\rangle &=&
(p_B+p_K)^\mu F_+(t) \nonumber \\
&+& (p_B-p_K)^\mu F_-(t) \\
 \langle \bar{K}^0(p_K)|\sbar b|\bar{B}^0_d(p_B)\rangle &=&
\delta_{bs}^{-1} \Delta_{BK} F_+(t) \nonumber \\
&+& \delta_{bs}^{-1} (p_B-p_K)^2 F_-(t) \\
 \langle \bar{K}^0(p_K)|\sbar \su(1\pm\gamma_5) b|\bar{B}^0_d(p_B)\rangle &=&
-2is(p_B^\mu p_K^\nu -p_B^\nu p_K^\mu) \nonumber \\
& \pm & s \varepsilon^{\mu\nu\lambda\sigma}
 (p_{B\lambda}p_{K\sigma}-p_{B\sigma}p_{K\lambda}).
\end{eqnarray}
Here $t=(p_B-p_k)^2$, $\Delta_{BK} \equiv m_{B}^{2}-m_{K}^{2}$, and 
$\delta_{bs} \equiv m_{b}-m_{s}$.
Notice we have corrected the sign in
Equation~(\ref{tensor}) compared 
to the similar expression in~\cite{Deshpande:1995tx}.
Heavy quark effective theory gives the relation
$s=(F_+-F_-)/4m_b$~\cite{Isgur:kf}. 
We also make the kinematic assumptions that the
$b$-quark carries all of the $B$-meson momentum
and that the $\phi$ momentum is equally divided between its two
constituent $s$-quarks. Thus $p_b=p_B$ and
$k^2=\frac{1}{2}(m_B^2-m_\phi^2/2+m_K^2)$. Putting all the pieces
together gives:
\begin{eqnarray}
\kappa &=& -\frac{m_b^2}{2k^2}\left[ 1+\frac{m_B^2-m_K^2+m_\phi^2
\frac{F_-}{F_+}}{4m_b(m_b-m_s)}\right] \nonumber \\
&+&-\frac{m_b^2}{2k^2}\left[\frac{m_K^2-m_B^2-2m_\phi^2}{8m_b^2}
\left(1-\frac{F_-}{F_+}\right)  + \frac{m_s}{2m_b} \left(
\frac{F_-}{F_+} \right) \right]
\end{eqnarray}
Because the matrix element is nonsingular we have $F_-(0)=0$. Then for
small $t$, due to simple pole dominance we have $F_-(m_\phi^2)\simeq
F_-(0) = 0$.~\cite{Cottingham:2001ax} By ignoring the difference
between $b$-quark and $B$-meson masses we arrive at the estimate cited
earlier, $\kappa =
-\frac{9}{8} + \mathcal{O}\left( \frac{m_\phi^2}{m_B^2} \right) 
\simeq -1.1$. The sign convention differs from~\cite{Barbieri:1997kq},
and the slight difference in magnitude can be traced to our replacement of
$(p_b+p_s)^\mu$ with $2p_b^\mu$ using the conservation of the added
quark current. In~\cite{Lunghi:2001af} a similar quantity
$\tilde{S}_{\phi K}=\frac{4}{9}\kappa$ is used. However, they quote a value
$\tilde{S}_{\phi K}\simeq -0.76$ which appears to match
$\tilde{S}_{\pi\pi}$ found in~\cite{Arhrib:2001gr}. This would correspond to a
value of $\kappa \simeq -1.71$.

%%%%%%%%%%%%%%%%%%%%%%%%%%%%%%%%%%%%%%%%%%%%%%%%%%%%%%%%%%%%%%%%%%%%%%%

\bibliographystyle{unsrt}

\end{document}